\documentclass[aps,prb,twocolumn,preprintnumbers]{revtex4}

\usepackage{amsmath}
\usepackage{graphicx}
\usepackage{soul}
\usepackage{amsfonts}
\usepackage{amssymb}
\usepackage{dcolumn}
\usepackage{adjustbox}
\usepackage{color}

\usepackage[T1]{fontenc}
\usepackage[latin9]{inputenc}
\usepackage{booktabs}

\usepackage{esint}
\usepackage{lipsum}

\usepackage{float}

\usepackage[utf8]{luainputenc}
\usepackage{rotating}
\usepackage{stackrel}
\usepackage{esint}

\makeatletter

\def\e{\begin{equation}}
\def\f{\end{equation}}
\def\=#1{\overline{\overline #1}}
\def\*#1{\overline{\overline{\overline #1}}}
\def\-#1{{\bf #1}}
\def\l#1{\label{eq:#1}}
\def\r#1{(\ref{eq:#1})}

\begin{document}

\title{Theory of metasurface based perfect absorbers}
\author{Rasoul Alaee,{$^{*1,2}$} Mohammad Albooyeh$^{3}$, Carsten Rockstuhl$^{1,4}$}
\affiliation{$^1$Institute of Theoretical Solid State Physics, Karlsruhe Institute of Technology, 76131 Karlsruhe, Germany\\
$^2$Max Planck Institute for the Science of Light, Erlangen 91058, Germany\\
$^3$Department of Electrical Engineering and Computer Science, University of California, Irvine, California 92697, USA\\
$^4$Institute of Nanotechnology, Karlsruhe Institute of Technology, 76021 Karlsruhe, Germany\\
~~*Corresponding author: rasoul.alaee@mpl.mpg.de~~}


\begin{abstract}
Based on an analytic approach, we present a theoretical review on the absorption, scattering, and extinction of both dipole scatterers and regular arrays composed of such scatterers i.e., metasurfaces. Besides offering a tutorial by outlining the maximum absorption limit for electrically/magnetically resonant dipole particles/metasurfaces, we give an educative analytical approach to their analysis. Moreover, we put forward the analysis of two known alternatives in providing perfect absorbers out of electrically and or magnetically resonant metasurfaces; one is based on the simultaneous presence of both electric and magnetic responses in so called Huygens metasurfaces while the other is established upon the presence of a back reflector in so called Salisbury absorbers. Our work is supported by several numerical examples to clarify the discussions in each stage.

\end{abstract}

\maketitle

\section{Introduction}

Light carries energy. In many applications, e.g. optical sensors,
thermal emitters, optical modulators, or solar cells, it is desirable
to absorb as much as possible the energy of the impinging light in some absorbing layers. This
requires to fully suppress the transmission and reflection while the
energy shall be dissipated in the absorbing layer. In principle, perfect
absorption occurs if two conditions are simultaneously satisfied: i.e., zero reflection and zero transmission. The first condition
requires the absorbing layer (which can be, e.g., an electrically resonant layer) to be impedance matched to the free
space. This fully suppresses the reflection. In order to
satisfy the second condition (i.e. zero transmission), there are several approaches. For example we may choose the absorbing
layer to be sufficiently thick (compared to the wavelength) or we may use several layers to create electromagnetic responses with both electric and magnetic properties.

\begin{figure}
   \centering
     \includegraphics[width=0.98\columnwidth]{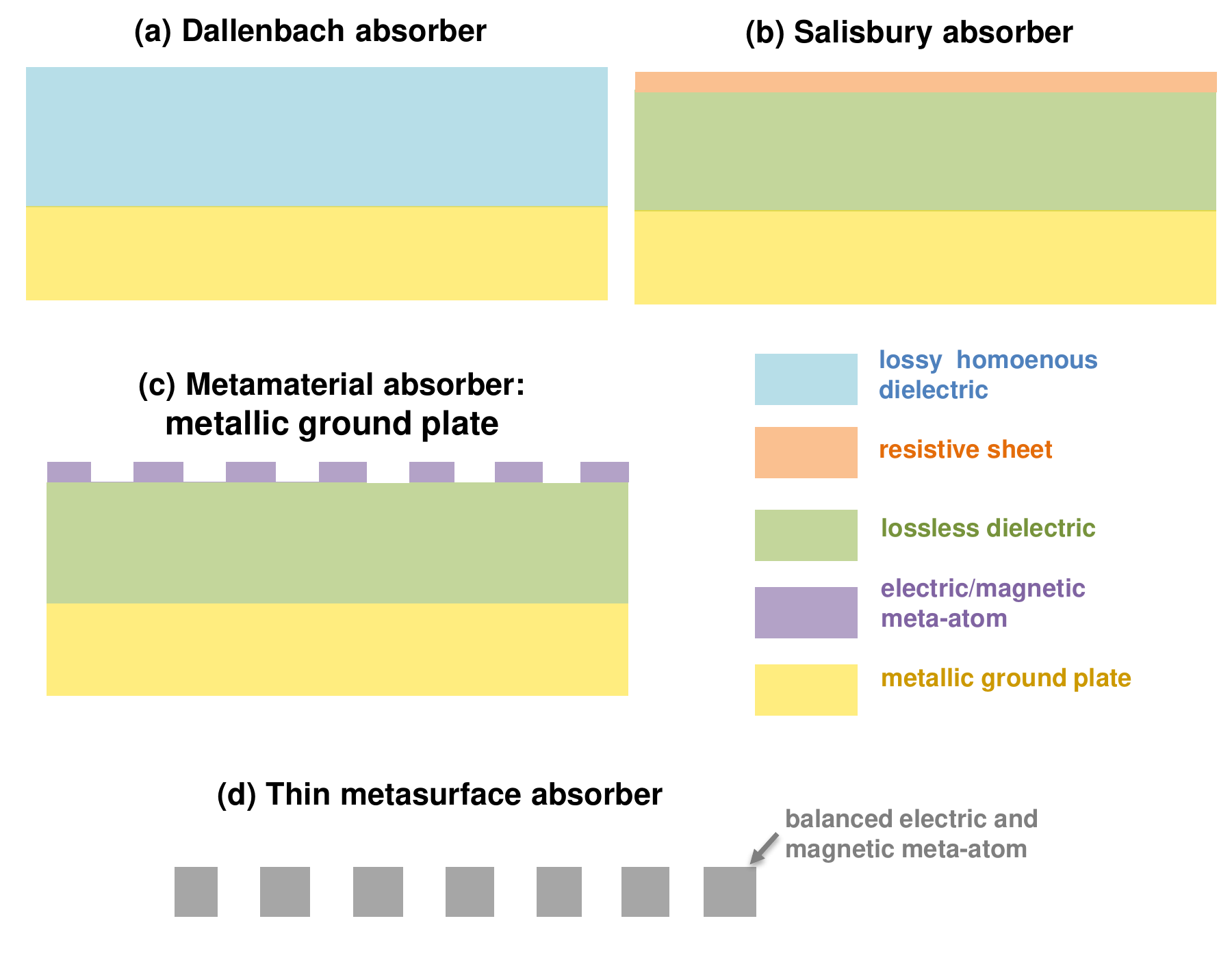}
\caption{Schematic view of three well-known perfect absorbers: (a) The Dallenbach
perfect absorber, (b) the Salisbury perfect absorber, (c) the metamaterial
perfect absorber made of electric/magnetic metasurface on top of a metallic ground plate, and (d) the thin metasurface perfect absorber made of balanced electric and magnetic meta-atoms.
\label{fig:Three_absorbers}}
\end{figure}

Research studies on how electromagnetic waves can be efficiently absorbed
potentially started with the invention of radar in the 1930's. At
that time, it was important to disguise an object (e.g. an airplane)
from radars. Therefore, various approaches have been introduced to
reduce the reflected waves from the object~\cite{Munk:05}. The most
important inventions among them are Dallenbach\cite{Dallenbach:38}
and Salisbury perfect absorbers~\cite{Salisbury:52}. Dallenbach
perfect absorbers consist of a homogeneous lossy dielectric layer
on top of a metallic ground plate~{[}Fig.~\ref{fig:Three_absorbers}
(a){]}. The thickness of the layer and its material properties
(permittivity and permeability) have to be carefully chosen such that
the reflected light is totally suppressed via impedance matching.
The thickness of the layer should be roughly a quarter wavelength.
In contrast, a Salisbury perfect absorber consists of a resistive
sheet on top of a metallic ground plate separated by a dielectric
spacer (no Ohmic losses)~{[}Fig.~\ref{fig:Three_absorbers} (b){]}.
This perfect absorber mainly relies on the absorption in the resistive
sheet. The refractive index of the dielectric spacer plays an important
role on choosing the thickness of the layer. The basic principle of
the Salisbury perfect absorbers can be understood by a destructive
interference of the reflected light and a suppression of the transmitted
light.

Recently, perfect metamaterial/metasurface absorbers~{[}Fig.~\ref{fig:Three_absorbers}
(c)-(d){]} have been introduced based on periodic subwavelength resonant particles (sometimes called meta-atoms or nanoantennas)\cite{Engheta:02,Landy:08}. Perfect metamaterial absorbers attracted considerable attentions at microwave~\cite{Landy:08,Watts:12,Radi:13}, terahertz, and especially near-infrared\cite{Liu:07,Avitzour:09,Liu:10,Pu:11,Piper:12,Li:12}, and visible\cite{Alaee_Absorber:12,Albooyeh:12,Aydin:11,Hu:09,Huebner:13} frequencies.

To understand the underlying physics of the metasurface based perfect absorbers, it
is very important to explore the scattering properties of an individual element of such metasurfaces, usually referred as
nanoantenna. For most pertinent nanoantennas the optical response
can be fully described by an electric and/or magnetic dipole moment. An important question in this regard, therefore, would be: Is there any universal limitation on the scattering, absorption, and extinction properties of a nanoantenna which exhibits electric/magnetic dipole response? Moreover, it is important to have an analytical approach available to predict the optical response of an array of nanoantennas based on the response of the individual
nanoantenna. By having this analytical approach, further questions one
might ask are: What is the maximal achievable absorption for a metasurface, which exhibit only an electric/magnetic response? Finally, is it possible to achieve complete light absorption for an array of dipole resonant scatterers? In this paper, we try to answer all the aforementioned questions by a theoretical review on the limits of absorption and scattering of both an individual dipole resonant scatterer and a condense array of such resonant scatterers i.e., metasurfaces~\cite{AlaeeThesis:15}. We support our theoretical review by adapted numerical simulations when required. Even though partially already documented in literature, these insights have not yet been presented in a concise manner. It is the purpose of this contribution to close this gap.

\section{Scattering properties of nanoantennas: Universal limitations}

Nanoantennas are canonical elements in nanooptics, which control light-matter interaction at the nanoscale. Nanoantennas
are usually made of two types of materials i.e. a) metals or b) dielectrics.
In the case of metals, nanoantennas support a localized surface plasmon polariton
resonance. This is a collective oscillation of the conduction electrons
of the metallic antenna resonantly coupled to the electromagnetic
wave at visible or near-infrared frequencies~\cite{Maier:07}. In
the case of dielectric nanoantennas, Mie resonances are excited, which are also known as whispering gallery resonances.
In this section, we want to review some universal limitations (in
the dipole approximation) on their interaction with light~\cite{Tretyakov:03,Pozar:09, Ruan:10,novotny:12,Tretyakov:14,Miller:16}.

\subsection{Electric dipole}
Let us start with the simplest case, i.e. a nanoantenna that only
supports an electric dipole moment $\mathbf{p}$. We start by considering an incident plane wave with an electric field $E_{x}^{\mathrm{inc}}$ polarized along
the $x$-axis that propagates in $z$-direction.
The induced electric dipole moment $p_{x}$ in the nanoantenna is expressed as~\cite{Tretyakov:03}
\[
p_{x}=\epsilon_{0}\alpha_{\textrm{ee}}E_{x}^{\textrm{inc}},
\]
where $\alpha_{\textrm{\textrm{e}e}}$ is the electric polarizability
(here, we assumed an
isotropic and electric dipolar particle so that $\alpha_{\mathrm{ee}}$ is a scalar) and $\epsilon_{0}$
is the free space permittivity. The scattered electric $\mathbf{E}_{\textrm{sca}}$ and magnetic $\mathbf{H}_{\textrm{sca}}$ far-fields by the induced electric dipole moment $p_{x}$ are given by~\cite{Jackson:98}\begin{eqnarray}
\mathbf{E}_{\textrm{sca}} & = & Z_{0}\mathbf{H}_{\textrm{sca}}\times\mathbf{n}=\frac{ck^{^{2}}}{4\pi}Z_{0}\left(\mathbf{n}\times\mathbf{p}\right)\times\mathbf{n}\frac{e^{ikr}}{r},\nonumber \\
\mathbf{H}_{\textrm{sca}} & = & \frac{ck^{^{2}}}{4\pi}\left(\mathbf{n}\times\mathbf{p}\right)\frac{e^{ikr}}{r},\label{eq:E_D}
\end{eqnarray}
where $\mathbf{n}$ is the normal unit vector along the radial direction of the Spherical Coordinates, $Z_0$ is the impedance
of the ambient material, $k=\omega/c$
is the wavenumber for an angular frequency $\omega$ in vacuum, and $c$ is the speed of light in vacuum.
The exponential dependence of the fields on the spatial coordinate and the frequency
is omitted here for brevity. By using the above expressions, the scattered
(or radiated) power (i.e. the surface integral of the outward flowing
flux of the Poynting vector of scattered fields) of the induced electric
dipole moment reads (see the Appendix)
\begin{eqnarray}
P_{\textrm{sca}} & = & \frac{1}{2}\textrm{Re}\oiint_{S}\left(\mathbf{E}_{\textrm{sca}}\times\mathbf{H}_{\textrm{sca}}^{^{*}}\right)\cdot\mathbf{n}\ \textrm{d}s\nonumber \\
 & = & \frac{c^{2}Z_{0}k^{4}}{12\pi}|\mathbf{p}|^{2}=\frac{k^{4}}{6\pi}|\alpha_{\textrm{ee}}|^{2}\frac{|E_{x}^{\textrm{inc}}|^{2}}{2Z_{0}}.\label{eq:P_scat}
\end{eqnarray}

Now by using the definition of the time-averaged Poynting vector for
the illuminating plane wave, by defining the intensity of the impinging light as $I_{\textrm{0}}=\frac{1}{2}\textrm{Re}\left(\mathbf{E}_{\textrm{inc}}\times\mathbf{H}_{\textrm{inc}}^{*}\right)=|E_{x}^{\textrm{inc}}|^{2}/2Z_{0}$,
and by using Eq.~\ref{eq:P_scat}, the total scattering cross section of the
induced electric dipole moment is expressed as

\begin{eqnarray}
C_{\textrm{sca}} & = & \frac{P_{\textrm{sca}}}{I_{0}}=\frac{k^{4}}{6\pi}|\alpha_{\textrm{ee}}|^{2}.\label{eq:C_sca_ED}
\end{eqnarray}

On the other hand, the extracted power is defined as~\cite{Huffman:98} (see the Appendix)

\begin{eqnarray}
P_{\textrm{ext}} & = & -\frac{1}{2}\textrm{Re}\oiint_{S}\left(\mathbf{E}_{\textrm{inc}}\times\mathbf{H}_{\textrm{sca}}^{*}+\mathbf{E}_{\textrm{sca}}\times\mathbf{H}_{\textrm{inc}}^{*}\right)\cdot\mathbf{n\,}ds\nonumber \\
 & = & -\frac{\omega}{2}\textrm{Im}\left(\mathbf{p}^{*}\cdot\mathbf{E}_{\textrm{inc}}\right)=\frac{1}{2}\omega\epsilon_{0}\alpha_{\textrm{ee}}^{\prime\prime}|\mathbf{E}_{\textrm{inc}}|^{2},\label{eq:Cext_ED}
\end{eqnarray}

where $\alpha_{\textrm{ee}}^{\prime\prime}$ is the imaginary part
of the electric polarizability. Similar to scattering, the extinction cross section is defined as $C_{\textrm{ext}}=\frac{P_{\textrm{ext}}}{I_{0}}=k\textrm{Im}\left(\alpha_{\textrm{ee}}\right)=k\alpha_{\textrm{ee}}^{\prime \prime}$.
It is well-known that the extracted power is the sum of the scattered power $P_{\textrm{sca}}$ and the absorbed
power $P_{\textrm{abs}}$ i.e., $P_{\textrm{ext}}=P_{\textrm{sca}}+P_{\textrm{abs}}$~\cite{Huffman:98}. The same is valid for
the cross sections, i.e. $C_{\textrm{ext}}=C_{\textrm{sca}}+C_{\textrm{abs}}$.
Therefore, the absorption cross section reads

\begin{eqnarray}
C_{\textrm{abs}} & = & \frac{P_{\textrm{abs}}}{I_{0}}=\frac{P_{\textrm{ext}}-P_{\textrm{sca}}}{I_{0}}\nonumber \\
 & = & k\alpha_{\textrm{ee}}^{\prime\prime}-\frac{k^{4}}{6\pi}|\alpha_{\textrm{ee}}|^{2}.\label{eq:C_abs}
\end{eqnarray}

Now we can calculate the maximum possible absorption cross section by differentiating
$C_{\textrm{abs}}$ with respect to the real and imaginary parts of
the polarizability, i.e. $\alpha_{\textrm{ee}}^{\prime}$ and $\alpha_{\textrm{ee}}^{\prime\prime}$ and the results read  $\alpha_{\textrm{ee}}^{\prime}=0,$~~$\alpha_{\textrm{ee}}^{\prime\prime}=3\pi/k^{3}$.
This suggests that in order to achieve a maximal absorption cross section,
the nanoantenna should be in resonance, i.e. the real part of the
polarizability should be zero. By substituting $\alpha_{\textrm{ee}}^{\prime}=0,$~~$\alpha_{\textrm{ee}}^{\prime\prime}=3\pi/k^{3}$
into Eq.~\ref{eq:C_abs}, the maximum absorption cross section reads $C_{\textrm{abs}}^{\textrm{max}}=\frac{3\pi}{2k^{2}}= \frac{3}{8\pi}\lambda^{2}$. At the same time, the
scattering cross section is $C_{\textrm{sca}}=\frac{3\pi}{2k^{2}}$. Therefore,
we can conclude that a nanoantenna which exhibits only an
electric dipole response is most absorptive if the absorbed power
is identical to its scattered power, i.e. $C_{\textrm{abs}}^{\textrm{max}}=C_{\textrm{sca}}=\frac{3\pi}{2k^{2}}.$
This condition is known as \textit{critical coupling}~\cite{Fan:11,Filonov:12,Alaee_Absorber:12,Tretyakov:14}.
Moreover, it is important to mention that for such a nanoantenna the
scattering/absorption cross section ($C_{\textrm{abs}}^{\textrm{max}}$
or $C_{\textrm{sca}}$) can exceed the geometrical cross section ($A$)
by an order of magnitude. To exemplify it, let us assume
a nanodisk antenna with a finite height, radius $a$ and geometrical
cross section of $A=\pi a^{2}$. The nanoantenna shall be sufficiently
small compared to the wavelength ($ka<<1$). Hence, the absorption/scattering
cross section can be much bigger than the geometrical cross section
at critical coupling, i.e.\e
\frac{C_{\textrm{sca}}}{A}=\frac{C_{\textrm{abs}}^{\textrm{max}}}{A} ={3\over {2(ka)^{2}}} ~ >>1.
\f
Note that for metallic nanoantennas the enhanced absorption is associated
with the excitation of surface plasmon polaritons~\cite{Maier:07}, which leads
to a huge near-field enhancement. This field enhancement will play
a crucial rule to achieve promising applications in quantum optics and optical sensing. For lossy dielectric
nanoantennas at infrared wavelengths the absorption is associated
with the excitation of phonons~\cite{Bohren:83}.

To gain further insights into the scattering response, let us study the implications of the analysis above to some limiting cases. For that, let us focus on a scatterer made of a non-absorbing material. It implies that the absorbed power is zero
($P_{\textrm{abs}}=0$). This means that the impinging energy of the
illuminating plane wave on the nanoantenna will be entirely scattered into
the surrounding media. Therefore, the extracted and scattered power
should be identical, i.e. $P_{\textrm{ext}}=P_{\textrm{sca}}$. This
equality leads to a well-known expression for the imaginary part of
the polarizability

\begin{eqnarray}
\textrm{Im}\left(\frac{1}{\alpha_{\textrm{ee}}}\right) & =- & \frac{\alpha_{\textrm{ee}}^{\prime\prime}}{|\alpha_{\textrm{ee}}|^{2}}=-\frac{k^{3}}{6\pi}.\label{eq:SL_ED}
\end{eqnarray}

Equation~\ref{eq:SL_ED} presents an important physical quantity
that is known as the \textit{scattering losses}; also called \textit{radiation
losses}~\cite{Sipe:74,Tretyakov:03}. Note that this quantity
only depends on the wavelength and is independent of the geometrical
parameters and materials that the nanoantenna is made of.

\begin{figure}
 \centering
  \includegraphics[width=0.98\columnwidth]{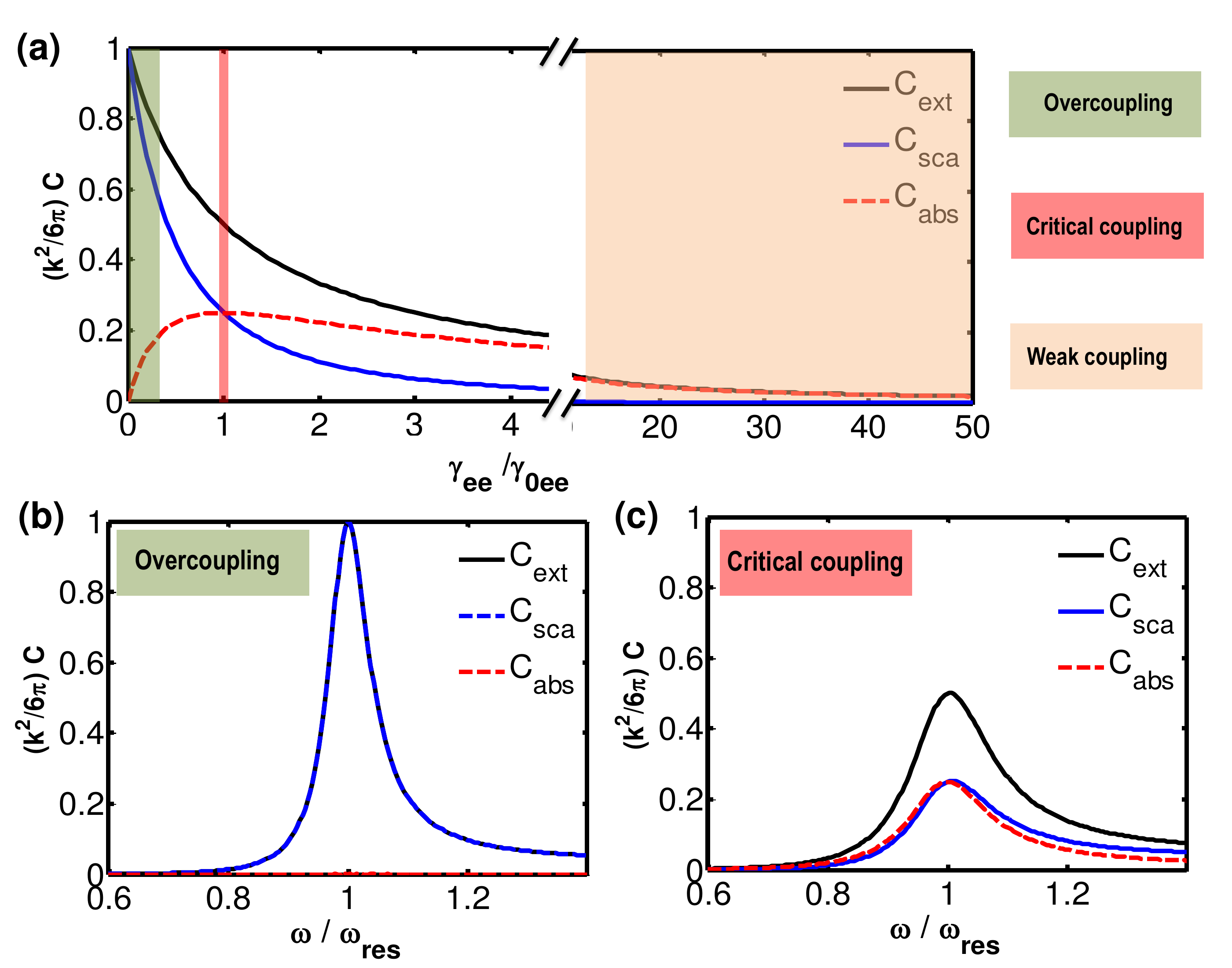}

\caption{(a) Normalized extinction, absorption, and scattering cross sections
at resonance as a function of the normalized Ohmic losses (i.e. $\gamma_{\textrm{ee}}/\gamma_{\textrm{ee0}}$).
(b) Normalized extinction, absorption, and scattering cross sections
as a function of the normalized frequency for the lossless case (i.e.
$\gamma_{\textrm{ee}}=0$). (c) The same plot as (b) but for the case
of maximum absorption cross section or critical coupling (i.e. $\gamma_{\textrm{ee}}=\gamma_{\textrm{ee}0}$).
\label{fig:SCS_Analy_maxim}}
\end{figure}
Next, let us assume that the nanoantenna is made of a lossy material. In order to contextualize the aforementioned results, let us assume
that the polarizability of the nanoantenna $\alpha_{\textrm{ee}}$
can be expressed by a Lorentzian line-shape that reads as~\cite{Sipe:74,Albooyeh:12}

\begin{eqnarray}
\alpha_{\textrm{ee}} & = & \frac{\alpha_{\textrm{0ee}}}{\omega_{\textrm{0ee}}^{2}-\omega^{2}-i\omega\gamma_{\textrm{ee}}-i\frac{k^{3}}{6\pi}\alpha_{\textrm{0ee}}},\label{eq:Alpha_ED}
\end{eqnarray}

where $\omega_{\textrm{0ee}}$ is the resonance frequency, $\gamma_{\textrm{ee}}$
is expressing the Ohmic losses (also called nonradiative losses), and $\alpha_{\textrm{0ee}}$
is the strength of resonance. The last term, i.e. $\frac{k^{3}}{6\pi}$,
is related to the scattering losses. Note that Eq.~\ref{eq:Alpha_ED}
fulfils Eq.~\ref{eq:SL_ED} if the Ohmic losses are zero, i.e. $\gamma_{\textrm{ee}}=0$.

By using Eqs.~\ref{eq:C_sca_ED},~\ref{eq:Cext_ED}, and~\ref{eq:SL_ED},
the scattering and extinction cross sections are given by

\begin{eqnarray}
C_{\textrm{sca}} & = & \frac{k^{4}}{6\pi}\frac{\alpha_{\textrm{0ee}}^{2}}{\left(\omega_{\textrm{0ee}}^{2}-\omega^{2}\right)^{2}+\left(\omega\gamma_{\textrm{ee}}+\frac{k^{3}}{6\pi}\alpha_{\textrm{0ee}}\right)^{2}},\label{eq:Csca_ED_}
\end{eqnarray}

\begin{eqnarray}
C_{\textrm{ext}} & = & \frac{k\alpha_{\textrm{0ee}}\left(\omega\gamma_{\textrm{ee}}+\frac{k^{3}}{6\pi}\alpha_{\textrm{0ee}}\right)}{\left(\omega_{\textrm{0ee}}^{2}-\omega^{2}\right)^{2}+\left(\omega\gamma_{\textrm{ee}}+\frac{k^{3}}{6\pi}\alpha_{\textrm{0ee}}\right)^{2}}.\label{eq:Cext_ED_}
\end{eqnarray}

\begin{figure*}
 \centering
  \includegraphics[width=0.7\textwidth]{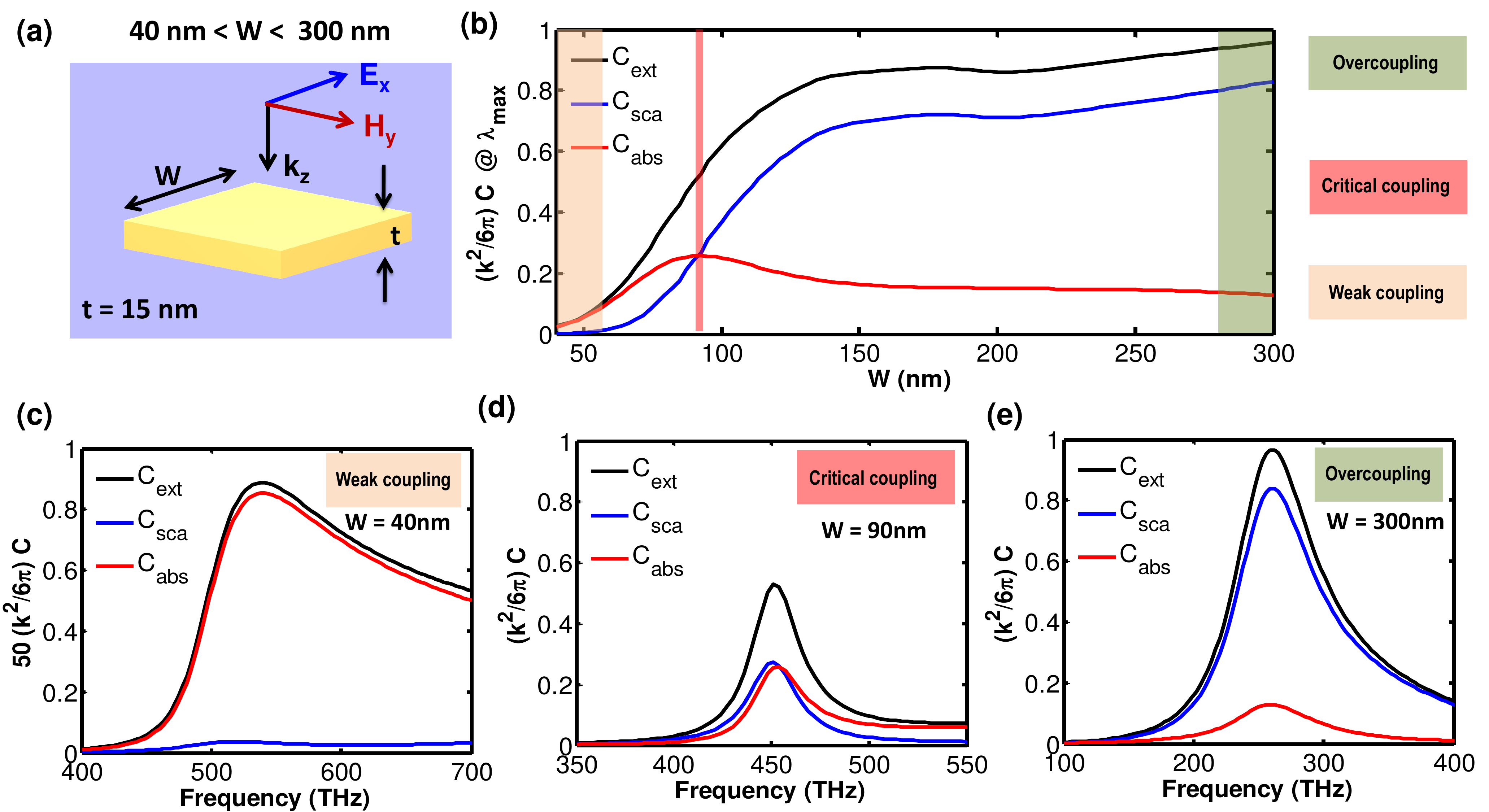}
\caption{(a) Schematic of the considered nanopatch antenna. (b) Normalized extinction,
scattering, absorption cross sections at resonance frequency ($\lambda_{\textrm{max}}$)
as a function of the width of the investigated nanopatch nanoantenna (i.e.
$W$). (c) Normalized extinction, scattering, absorption cross sections
as a function of frequency when the width of the nanoantenna is $W=40$~nm.
In this case the $C_{\textrm{ext}}\approx C_{\textrm{abs}}$. This
occurs for nanoantennas really small compared to the wavelength. (d)
Same as (c) whenever the nanoantenna is at critical coupling, i.e.
$C_{\textrm{abs}}|_{\lambda_{\textrm{max}}}=C_{\textrm{ext}}|_{\lambda_{\textrm{max}}}$.
This occurs if the width of nanoantenna is $W=90$~nm. (e) Same as
(c) when the scattering cross section dominates, here at $W=300$~nm. Note that by using induced electric current, it is easy to show that the considered nanopatch antenna only supports an electric dipole moment~\cite{Alaee:18}.
\label{fig:SinglePatch_Ultimate}}
\end{figure*}


We calculated the scattering, absorption, and extinction cross sections
as a function of the normalized Ohmic losses at the resonance frequency
{[}Fig.~\ref{fig:SCS_Analy_maxim}~(a){]}. They are normalized to $\frac{6\pi}{k^{2}}$,
which is the maximum extinction cross section. Moreover, the Ohmic losses are normalized to the scattering losses. In principle, three different
coupling regimes are distinguished:

a) \textit{Overcoupling} ($\gamma_{\textrm{ee}}<<\gamma_{0\textrm{ee}}$):
This occurs if the Ohmic losses (i.e. $\gamma_{\textrm{ee}}$) of
the nanoantenna are much smaller than the scattering losses (i.e.
$\gamma_{\textrm{0ee}}$). The scattering losses are defined
as $\gamma_{0\textrm{ee}}=\frac{k^{3}}{6\pi}\left(\frac{\alpha_{\textrm{0ee}}}{\omega_{0}}\right)$.
Figure~\ref{fig:SCS_Analy_maxim} (b) shows the normalized scattering,
absorption, and extinction cross sections as a function of the normalized
frequency (i.e. $\omega/\omega_{\textrm{res}}$) for the lossless
case, i.e. $\gamma_{\textrm{ee}}=0$. It can be seen that the scattering
and extinction cross sections are identical, i.e. $C_{\textrm{ext}}=C_{\textrm{sca}}$.

b) \textit{Critical coupling} ($\gamma_{\textrm{ee}}=\gamma_{0\textrm{ee}}$):
This occurs whenever the Ohmic losses ($\gamma_{\textrm{ee}}$) are
identical to the scattering losses ($\gamma_{\textrm{0ee}}$). This
condition allows to achieve maximum absorption. It can be seen that
the maximum absorption cross section occurs whenever $\gamma_{\textrm{ee}}=\gamma_{0\textrm{ee}}=\frac{k^{3}}{6\pi}\left(\frac{\alpha_{\textrm{0ee}}}{\omega_{0}}\right)$
{[}Fig.~\ref{fig:SCS_Analy_maxim} (a) and (c){]}. Figure~\ref{fig:SCS_Analy_maxim}
(c) shows the same plot as Fig.~\ref{fig:SCS_Analy_maxim} (b) for
the case of critical coupling. In this case, the absorption cross
section is identical to the scattering cross section, i.e. $C_{\textrm{abs}}=C_{\textrm{sca}}$.
Therefore, the extinction cross section will be $C_{\textrm{ext}}=2C_{\textrm{abs}}=2C_{\textrm{sca}}$.

c) \textit{Weak coupling} ($\gamma_{\textrm{ee}}>>\gamma_{0\textrm{ee}}$):
This occurs if the Ohmic losses ($\gamma_{\textrm{ee}}$) are much
larger than the scattering losses ($\gamma_{\textrm{0ee}}$). At extremely
large Ohmic losses, i.e. $\gamma_{\textrm{ee}}>>\gamma_{0\textrm{ee}}$,
the scattering cross section is considerably small with respect to
the absorption cross section, i.e. $C_{\textrm{sca}}<<C_{\textrm{abs}}$.
In other words, the absorption cross section is approximately identical to the extinction
cross section, i.e. $C_{\textrm{ext}}\simeq C_{\textrm{abs}}$. This
occurs whenever the nanoantenna is considerably small with respect
to the wavelength or the dissipative losses of nanoantenna material is too high. The frequency dependency of the scattering, absorption,
and extinction cross sections for the weak coupling regime is not
shown here.

The aforementioned theoretical results can be applied to an arbitrary
plasmonic nanoantenna as long as it supports \textit{only} an electric
dipole response. In order to confirm that, we investigate possibly
the simplest nanoantenna, i.e. a gold nanopatch {[}Fig.~\ref{fig:SinglePatch_Ultimate}~(a){]}. The permittivity of gold is taken from Ref.~\onlinecite{Johnson:72}. We used a numerical finite element solver to obtain all the numerical results~\cite{multiphysics2012}. The height of the nanopatch is assumed to be $t=15$~nm and its width
$W$ is varying from $40$~nm to $300$~nm. It can be seen that
by varying the width $W$ of the nanoantenna, it is possible to tune
the Ohmic losses. This is shown in Fig\@.~\ref{fig:SinglePatch_Ultimate}~(a).
For small nanoantennas, i.e. $W<60$~nm, the absorption cross section
is much larger than the scattering cross section, i.e. $C_{\textrm{abs}}\simeq C_{\textrm{ext}}$
{[}Fig\@.~\ref{fig:SinglePatch_Ultimate} (c){]}. In principle,
this is always valid for a nanoantenna much smaller than the operating
wavelength. By increasing the width of the nanopatch, the condition for critical
coupling (maximum absorption regime) is met. In the chosen example
this happens for $W=90$~nm {[}Fig\@.~\ref{fig:SinglePatch_Ultimate}(c){]}.
Finally, for considerably large nanoantenna, scattering losses dominate
and the extinction cross section might reach its maximum value, i.e.
$C_{\textrm{ext}}=\frac{6\pi}{k^{2}}=\frac{3\lambda^2}{2\pi}$. In other words, the
absorption cross section is small compared to the scattering cross
section, i.e. $C_{\textrm{ext}}\simeq C_{\textrm{sca}}>>C_{\textrm{abs}}$.

\begin{figure}
\includegraphics[width=0.98\columnwidth]{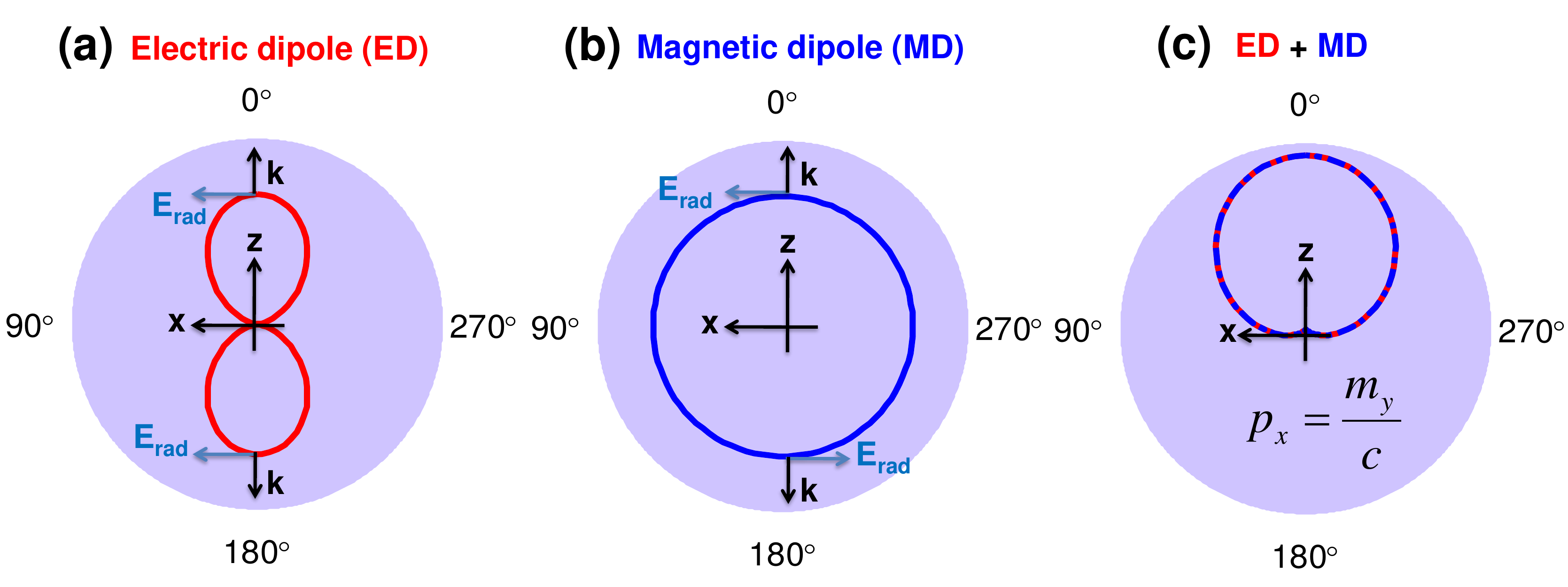}
\caption{Radiation pattern for different multipole moments in the xz-plane.
(a) Electric dipole moment ($\textrm{p}_{x}$), i.e. $|\mathbf{E}_{\textrm{far}}|^{2}\propto|p_{\textrm{x}}|^{2}\cos^{2}\theta$.
The blue arrows indicate the phase of the radiated field. (b) Magnetic
dipole moment ($m{}_{\textrm{y}}$), i.e. $|\mathbf{E}_{\textrm{far}}|^{2}\propto|\frac{m_{y}}{c}|^{2}$.
(c) Superposition of electric and magnetic dipole moments, i.e. $|\mathbf{E}_{\textrm{far}}|^{2}\propto|p_{\textrm{x}}|^{2}(1+\cos\theta)^{2}$,
when the Kerker condition is fulfilled, i.e. $p_{\textrm{x}}=\frac{m_{\textrm{y}}}{c}$.
\label{fig:Directivity_Radiation-pattern_ED_MD}}
\end{figure}

All the results in this subsection are equally valid when considering a nanoantenna
that supports a magnetic dipole moment (see e.g. Refs.~\onlinecite{Darvishzadeh:17,Alaee:15,AlbooyehAbsorber}). All the equations for
a magnetic dipole moment are analogous, or dual, to those for an electric
dipole moment. In fact, this can be understood from the duality of the electric and magnetic fields/currents in Maxwell's equations. A similar expression to that of Eq.~\ref{eq:SL_ED} also holds for the scattering losses of a magnetic dipole, i.e. $\textrm{Im}\left(\frac{1}{\alpha_{\textrm{mm}}}\right)=-\frac{k^{3}}{6\pi}$.
This expression is called Sipe-Kranendonk condition that can
be obtained from conservation of energy~\cite{Sipe:74,Tretyakov:03,Swiecicki2017}.
Note that the above discussions are valid as long as the nanoantenna exhibits
only an electric/magnetic dipole moment. It is also interesting to
mention that these universal limitations are independent of the shape
and material of the nanoantenna. Experimental verification of some
of the above results can be found in~\cite{Husnik:08,Husnik:12,Husnik:13}.

\subsection{Electric and magnetic dipoles and higher order multipoles}
For a dipolar particle that simultaneously supports an electric and magnetic response represented, respectively, by electric and magnetic dipole moments $p_{x}$
and $m_{y}$, the radiated electric far-field reads~\cite{Jackson:98,Alaee_phase:16}

\begin{eqnarray}
\mathbf{E}_{\mathrm{sca}}\left(\mathbf{r}\right) & = & \frac{k^{^{2}}}{4\pi\epsilon_{0}}p_{x}\frac{e^{ikr}}{r}\left(-\mathrm{sin}\varphi\mathbf{\hat{\varphi}}+\mathrm{cos}\theta\mathrm{cos}\varphi\hat{\theta}\right)\label{eq:Efar_ED_MD}\\
 &  & -\frac{k^{^{2}}}{4\pi\epsilon_{0}}\frac{m_{y}}{c}\frac{e^{ikr}}{r}\left(\mathrm{cos}\theta\mathrm{sin}\varphi\mathbf{\hat{\varphi}}-\mathrm{cos}\varphi\hat{\mathbf{\theta}}\right),\nonumber
\end{eqnarray}

where, $r$ is the distance to the observation point, $\phi$ is the azimuthal angle, and $\theta$ is the polar angle in spherical coordinates. The extinction and scattering cross sections, respectively, read\begin{eqnarray}
C_{\textrm{ext}} & = & C_{\textrm{ext}}^{p}+C_{\textrm{ext}}^{m}\\
 & = & k\textrm{Im}\left(\alpha_{\textrm{ee}}+\alpha_{\textrm{mm}}\right),\nonumber \\
C_{\textrm{sca}} & = & \frac{k^{4}}{6\pi}\left(|\alpha_{\textrm{ee}}|^{2}+|\alpha_{\textrm{mm}}|^{2}\right).
\end{eqnarray}

The maximum scattering cross section is $C_{\textrm{ext}}=2\frac{3\lambda^{2}}{2\pi}$,
which is two times larger than that of an individual electric or a magnetic dipole. This occurs if the antenna is in the overcoupling regime and the Kerker condition $p_{x}=m_{y}/c$ is fulfilled~($C_{\textrm{ext}}^{p}=C_{\textrm{ext}}^{m}$). To illustrate the physical mechanism behind the Kerker condition,
the radiation pattern of such dipole antenna in the xz-plane ($\varphi=0$) is
shown in Figs.~\ref{fig:Directivity_Radiation-pattern_ED_MD}~(a)
and (b). The electric field radiated by an electric dipole and a magnetic
dipole interfere constructively at $\theta=0$ {[}Fig.~\ref{fig:Directivity_Radiation-pattern_ED_MD}~(c){]}, i.e. their radiated fields are in-phase in forward direction {[}Fig.~\ref{fig:Directivity_Radiation-pattern_ED_MD}~(a) and (b){]}. On the other hand, there is no scattering in backward direction ($\theta=\pi$) because of the destructive interference
in this direction. The maximum absorption cross section $C_{\textrm{abs}}=2\frac{3}{8\pi}\lambda^{2}$ occurs at critical coupling.

For particles large compared to the wavelength, higher order multipole moments (e.g. quadrupole, octupole, etc.) will be excited. By applying the energy conservation principle, one can show that the maximum scattering/extinction cross section of an isotropic particle is $\left(2j+1\right)\lambda^{2}/2\pi$~\cite{Ruan:10,Fan:11,Aso:17}, where $j$ is the total angular momentum number and $j=1,2,3$, respectively, correspond to the electric/magnetic dipole, quadrupole, and octupole moments. Similarly, the maximum absorption cross section is $\left(2j+1\right)\lambda^{2}/8\pi$~\cite{Ruan:10,Fan:11,Aso:17}. These fundamental limits for the scattering, absorption and extinction cross sections lead to some upper bounds for optical force and torque~\cite{Aso:17}.

\section{Optical properties of an array of electric nanoantennas}

The interaction of light with a periodic array of optically small resonant nanoantennas (also called metasurfaces) is an interesting and fundamental
problem which attracts tremendous interests~\cite{Abajo:07,Halas:11,Menzel:12}. In this section,
we concisely study and present a summary on the limitation concerning the optical properties
of such an array of nanoantennas. We assume that nanoantennas are
optically small and the externally incidence field only excites
electric dipole moments of nanoantennas. We show that a periodic array of electric
dipoles maximally absorbs $50$ percent of the impinging light.

\begin{figure}
 \centering
  \includegraphics[width=0.8\columnwidth]{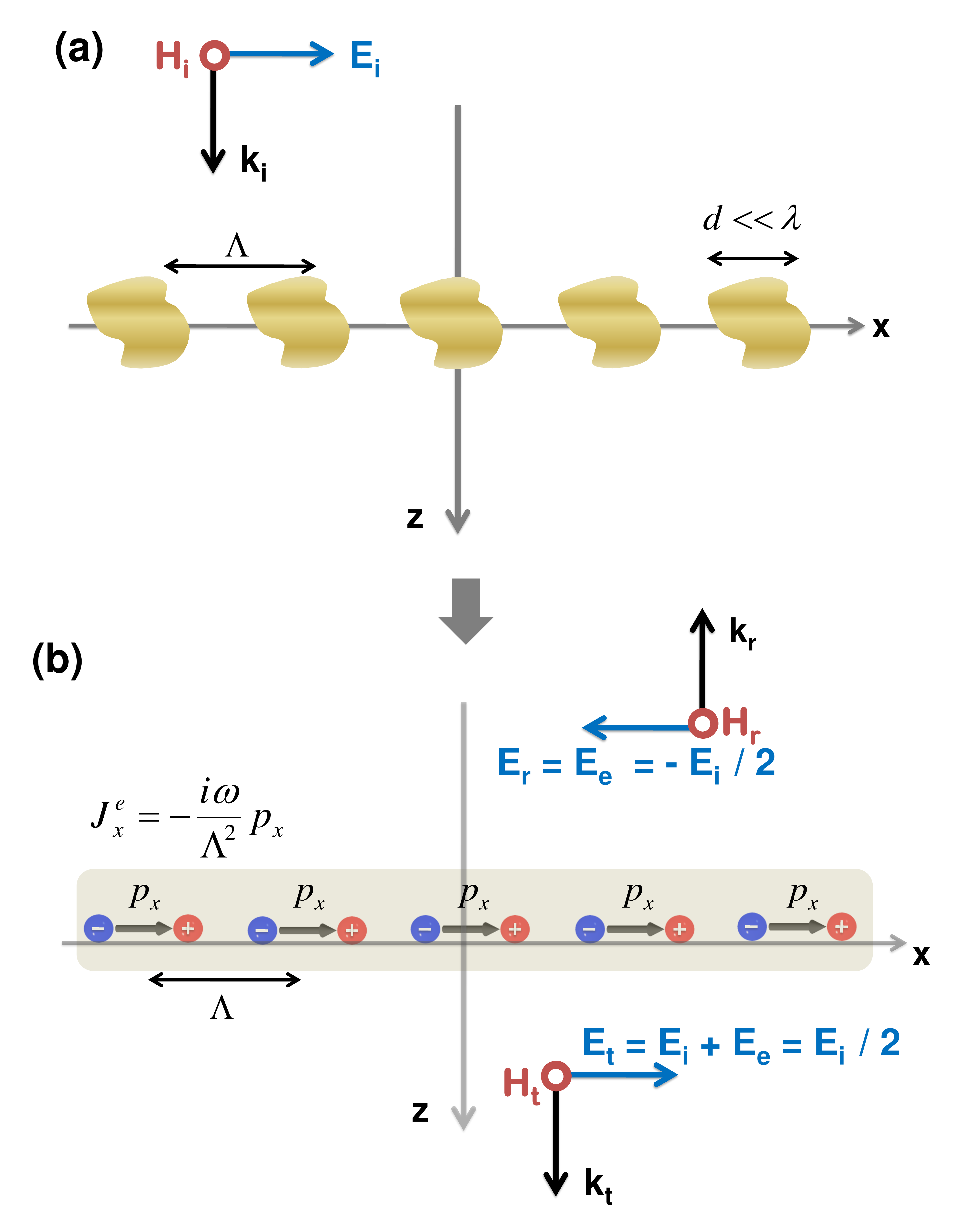}
\caption{(a) Geometry of an infinite periodic array of nanoantennas that support
only an electric dipole moment. The array is embedded in a uniform homogeneous
host medium and excited by a plane wave. The two-dimensional array
is periodic in the x-y plane. (b) The radiated fields by the average electric
current density $J_{x}$ in the case of maximum absorption,
i.e. $A_{\textrm{max}}=0.5$. \label{fig:UL_dipole}}
\end{figure}

Let us consider an infinite planar array of optically small resonant nanoantennas,
which support only an electric dipole moment. The array is excited by a plane wave
(normal incidence) with an electric field polarized along the $x$-axis
that propagates in the positive $z$-direction {[}Fig.~\ref{fig:UL_dipole}{]} so each dipole must be induced along the x-axis.
We assume that the array of nanoantennas is situated in the xy-plane
and embedded in a homogeneous lossless medium. The period $\Lambda$
of the array is sufficiently small compared to the wavelength and $W<<\lambda$ where
$W$ is a measure for the spatial extent of the individual nanoantenna.
The shape of the nanoantennas can be arbitrary, e.g. nanopatches,
nanodiscs, nanostrips, nanospheres, or nanocylinders. The only requirement is to have a nanoantenna topology to induce only a co-polarized electric dipole moment,
i.e. $p_{x}$ (without cross polarization), along the polarization of the incidence. All the higher order multipole moments should
be negligible. Now, in order to calculate the optical response of
a periodic array of electric dipoles {[}Fig.~\ref{fig:UL_dipole}
(b){]}, let us express the induced electric dipole moment $p_{x}$.
It reads comparable to that of a sample nanoantenna as
\begin{eqnarray}
p_{x} & = & \epsilon_{0}\alpha_{\textrm{ee}}E_{x}^{\textrm{loc}},\label{eq:P_ED}
\end{eqnarray}

where $E_{x}^{\textrm{loc}}$
is the local electric field at the dipole position and is defined
as $E_{x}^{\textrm{loc}}=E_{x}^{\textrm{inc}}+E_{x}^{\textrm{int}}$.
$E_{x}^{\textrm{inc}}$ is the incident electric field and
$E_{x}^{\textrm{int}}$ is the interaction field created
by the rest of the electric dipoles in the array at the position of the sample dipole~\cite{Tretyakov:03}.
The interaction field is proportional to $p_{x}$ and can
be expressed by $E_{x}^{\textrm{int}}=\frac{\beta_{\textrm{ee}}p_{x}}{\epsilon_{0}}$.
Here, the interaction constant $\beta_{\textrm{ee}}$ is given by~\cite{Tretyakov:03,Collin:60,yazdi}
\begin{eqnarray}
\beta_{\mathrm{ee}} & = & \frac{ik}{4\Lambda^{2}}\left(1+\frac{1}{ikR_{0}}\right)e^{ikR_{0}},\label{eq:ED_Inter}
\end{eqnarray}
where $R_{0}=\Lambda/1.438$ is the effective inter-particle distance.
This is calculated for the quasi-static case and
it is a good approximation as long as the array is condense and the nanoantennas are small compared
to the wavelength. Details about the derivation of the interaction
constant $\beta_{\textrm{ee}}$ can be found in Refs.~\onlinecite{Tretyakov:03,Collin:60,yazdi}.
Note that the imaginary part of the interaction constant for a periodic
array can be calculated without any approximation by using a conservation
of energy~\cite{Tretyakov:03} and can be written as
\begin{eqnarray}
\textrm{Im}\left(\beta_{\mathrm{ee}}\right) & = & \textrm{Im}\left(\frac{1}{\alpha_{\textrm{ee}}}\right)+\frac{k}{2\Lambda^{2}}=-\frac{k^{3}}{6\pi}+\frac{k}{2\Lambda^{2}}.\label{inconst}
\end{eqnarray}

It is very important to notice that the first term in Eq.~\ref{inconst} cancels out with the scattering losses (i.e. $-\frac{k^{3}}{6\pi}$) in regular arrays
and the second term (i.e. $\frac{k}{2\Lambda^{2}}$) is related to
the plane wave contribution created by the induced averaged electric surface current.
Finally, the interaction constant reads~\cite{Tretyakov:03}
\begin{eqnarray}
\beta_{\textrm{ee}} & = & \textrm{Re}\left(\beta_{\textrm{ee}}\right)+i\textrm{Im}\left(\beta_{\textrm{ee}}\right) \\
 & = & \textrm{Re}\left[\frac{ik}{4\Lambda^{2}}\left(1+\frac{1}{ikR_{0}}\right)e^{ikR_{0}}\right]+i\left(-\frac{k^{3}}{6\pi}+\frac{k}{2\Lambda^{2}}\right).\nonumber
\end{eqnarray}

By using Eq.~\ref{eq:P_ED} and the local field definition, the induced
electric dipole moment reads~\cite{Tretyakov:03}
\begin{eqnarray}
p_{x} & = & \epsilon_{0}\frac{\alpha_{\textrm{ee}}}{1-\alpha_{\textrm{ee}}\beta_{\textrm{ee}}}E_{x}^{\textrm{inc}}=\epsilon_{0}\alpha_{\textrm{eff}}E_{x}^{\textrm{inc}},\label{eq:px_dipole}
\end{eqnarray}

where $\alpha_{\textrm{ee}}^{\textrm{eff}}=\alpha_{\textrm{ee}}/\left(1-\alpha_{\textrm{ee}}\beta_{\textrm{ee}}\right)$
is the so-called effective or renormalized electric polarizability
of the array. To calculate the reflection and transmission coefficients,
we use the generalized boundary condition~\cite{newIdem, Holloway:05,Tretyakov:03,Albooyeh:2015,MohammadA},
i.e. $E_{x}^{\textrm{r}}=-\frac{1}{2}Z_0J_{x}$ and
the induced averaged electric surface current density, i.e. $J_{x}^{\textrm{e}}=-i\omega\frac{p_{x}}{\Lambda^{2}}$.
Note that the average magnetic surface current is zero. Therefore,
the reflected electric field is given as
\begin{eqnarray}
E_{x}^{\textrm{r}} & = & -\frac{1}{2}Z_{0}J_{x}^{\textrm{e}}=\frac{i\omega}{2\Lambda^{2}}Z_{0}p_{x}\nonumber \\
 & = & \frac{ik}{2\Lambda^{2}}\frac{\alpha_{\textrm{ee}}}{1-\alpha_{\textrm{ee}}\beta_{\textrm{ee}}}E_{x}^{\textrm{inc}},
\end{eqnarray}

and finally the reflection and transmission coefficients reads
\begin{eqnarray}
r & = & \frac{E_{x}^{\textrm{r}}}{E_{x}^{\textrm{inc}}}=\frac{ik}{2\Lambda^{2}}\frac{\alpha_{\textrm{ee}}}{1-\alpha_{\textrm{ee}}\beta_{\textrm{ee}}},\label{eq:R_ED}\\
t & = & \frac{E_{x}^{\textrm{t}}}{E_{x}^{\textrm{inc}}}=1+\frac{ik}{2\Lambda^{2}}\frac{\alpha_{\textrm{ee}}}{1-\alpha_{\textrm{ee}}\beta_{\textrm{ee}}}.\label{eq:T_ED}
\end{eqnarray}

Equations~\ref{eq:R_ED} and \ref{eq:T_ED} hold only for
an optically thin layer of nanoantennas. The absorption is calculated from the reflection and transmission coefficients and reads $A=1-|r|^{2}-|1+r|^{2}.$ The reflection coefficient is a complex number and can be written as $r=r_{r}+ir_{i}.$ Here, $r_{r}$
is the real part and $r_{i}$ is the imaginary part of the complex
reflection coefficient $r$. Thus, the absorption in terms of real and imaginary parts of the reflection coefficient can be written as
$A=-2\left(1+r_{r}\right)r_{r}-2r_{i}^{2}.$ Now, we can find the
condition for maximum absorption for an array that supports only electric dipole moments
by differentiating the absorption $A$ with respect to $r_{r}$ and
$r_{i}$ and searching for the roots, identifying extremal points
in the absorption, i.e. $r_{r}=-\frac{1}{2}$, $r_{i}=0.$ From that analysis it turns out that
in order to obtain maximum absorption, the reflection and transmission
coefficients of the array should be $r=-\frac{1}{2},\,\,t=1+r=\frac{1}{2}$.
By substituting these values of $r$ and $t$ into $A=1-|r|^{2}-|1+r|^{2}$,
one can see that an array of resonant electric dipoles maximally absorbs $50$ percent of the impinging light, i.e. $A_{\textrm{max}}=\frac{1}{2}$.
This is a universal limitation that holds for periodic arrays as long as it supports only an electric response. In order to cancel
the incident field, the array of electric dipoles should radiate
a plane wave with the same amplitude in both forward and backward
directions. The optimum case occurs when only half of the incident
field is canceled as shown in Fig.~\ref{fig:UL_dipole}. In the case
of the maximum absorption, the induced averaged electric surface current
density will be $J_{x}^{\textrm{e}}=\frac{E_{x}^{\textrm{inc}}}{Z}$. The corresponding individual polarizability is given by
\begin{eqnarray}
\alpha_{\textrm{ee}} & = & \frac{1}{\beta_{\textrm{ee}}-\frac{ik}{\Lambda^{2}}}.\label{eq:Alpha_ED_A_Max}
\end{eqnarray}

For a dense array of lossless (no Ohmic losses) nanoantennas, it can
be easily shown that the reflection coefficient at resonance will
reach $r=-1$. Therefore, the transmission coefficient necessarily has to be zero
$t=1+r=0$. As a result, the corresponding individual polarizability
can be expressed as
\begin{eqnarray}
\alpha_{\textrm{ee}} & = & \frac{1}{\beta_{\textrm{ee}}-\frac{ik}{2\Lambda^{2}}}.
\end{eqnarray}

In order to quantitatively understand the fundamental limitation of
absorption for an array of electric dipole moments, let us now assume
that the polarizability of an individual nanoantenna possesses a Lorentzian
line-shape given by Eq.~\ref{eq:Alpha_ED}. By using Eq.~\ref{eq:R_ED}
and the inter-particle interaction constant, i.e. $\beta_{\textrm{ee}}=\textrm{Re}\left(\beta_{\textrm{ee}}\right)-i\frac{k^{3}}{6\pi}+i\frac{k}{2\Lambda^{2}}$~\cite{Tretyakov:03},
the reflection coefficient reads
\begin{eqnarray}
r & = & \frac{ik}{2\Lambda^{2}}\frac{\alpha_{\textrm{0ee}}}{\omega_{0}^{'2}-\omega^{2}-i\omega\left(\gamma_{\textrm{ee}}+\frac{\alpha_{0}}{2\Lambda^{2}c_{0}}\right)},
\end{eqnarray}

where $\omega_{0}^{'}=\omega_{0}-\textrm{Re}\left(\beta_{\textrm{ee}}\right)\alpha_{\textrm{0ee}}$
is the resonance frequency of the array, which is shifted compared
to the resonance of an individual nanoantenna~(i.e.~$\omega_{0}$).
The reflection coefficient at the resonance frequency $\omega=\omega_{0}^{'}$
is given by
\begin{equation}
r\left(\omega=\omega_{0}^{'}\right)  =  -\frac{\frac{\alpha_{0}}{2\Lambda^{2}c_{0}}}{\left(\gamma_{\textrm{ee}}+\frac{\alpha_{0}}{2\Lambda^{2}c_{0}}\right)}=-\frac{\gamma_{\textrm{0ee}}}{\left(\gamma_{\textrm{ee}}+\gamma_{\textrm{0ee}}\right)},
\end{equation}
where $\gamma_{\textrm{0ee}}=\frac{\alpha_{0}}{2\Lambda^{2}c_{0}}$  while the transmission coefficient will be $t\left(\omega=\omega_{0}^{'}\right)=1+r\left(\omega=\omega_{0}^{'}\right)= {\gamma_{\textrm{ee}}}/{\left(\gamma_{\textrm{ee}}+\gamma_{\textrm{0ee}}\right)}$.

\begin{figure}
 \centering
  \includegraphics[width=0.98\columnwidth]{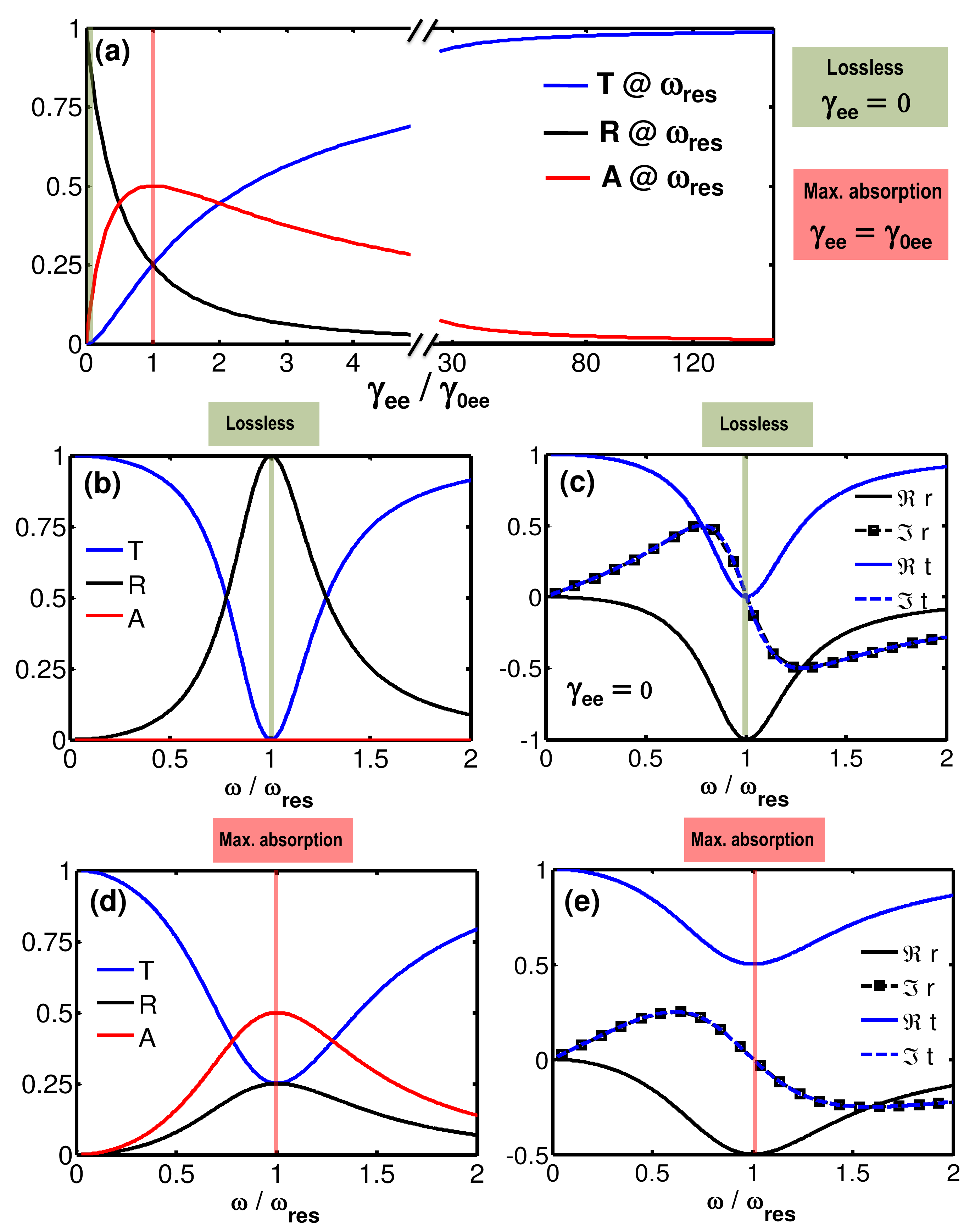}
\caption{(a) Transmission, reflection, and absorption at resonance frequency
as a function of normalized dissipation losses (i.e.~$\gamma_{\textrm{ee}}/\gamma_{\textrm{0ee}}$
). Maximum absorption (i.e.~$A_{\textrm{max}}=0.5$) occurs whenever
$\gamma_{\textrm{ee}}=\gamma_{\textrm{0ee}}$ (red line). (b) The
transmission, reflection, and absorption as a function of normalized
frequency for the lossless case, i.e. $\gamma_{\textrm{ee}}=0.$ (c)
The same plot as (b) for the reflection and transmission coefficients.
(d) -(e) Same plots as (b) and (c) for the case of maximal absorption
(i.e.~$\gamma_{\textrm{ee}}=\gamma_{\textrm{0ee}}$). \label{fig:-UniLimitation_EDipole}}
\end{figure}

To better understand all the aforementioned results, the reflection,
transmission, and absorption of an array of electric dipoles at the resonance frequency as a function of the normalized dissipation
losses (i.e.~$\gamma_{\textrm{ee}}/\gamma_{\textrm{ee0}}$) are shown
in Fig.~\ref{fig:-UniLimitation_EDipole} (a). It can be seen that
the maximum reflection (i.e. $R_{\textrm{max}}=|r_{\textrm{max}}|^{2}=1$,
$r_{\textrm{max}}=-1$) occurs if the nanoantennas in the array are
lossless, i.e. $\gamma_{\textrm{ee}}=0$ {[}Fig.~\ref{fig:-UniLimitation_EDipole}
(a)-(c){]}. The transmission ($t=1+r$) for such a lossless array
goes to zero at resonance frequency {[}Fig.~\ref{fig:-UniLimitation_EDipole}~(b)
and (c){]}. The array achieves its maximum absorption, i.e. $A_{\textrm{max}}=\frac{1}{2}$,
if the dissipation loss is equal to $\gamma_{\textrm{ee}}=\gamma_{\textrm{0ee}}=\frac{\alpha_{0}}{2\Lambda^{2}c_{0}}$.
This point of operation is shown with red line in Fig.~\ref{fig:-UniLimitation_EDipole}
(a) and (d). In this case, the reflection and transmission coefficients
will be $r=-1/2$ ans $t=1/2$, which can easily be seen in Fig.~\ref{fig:-UniLimitation_EDipole}~(e).

\begin{figure}
 \centering
  \includegraphics[width=0.98\columnwidth]{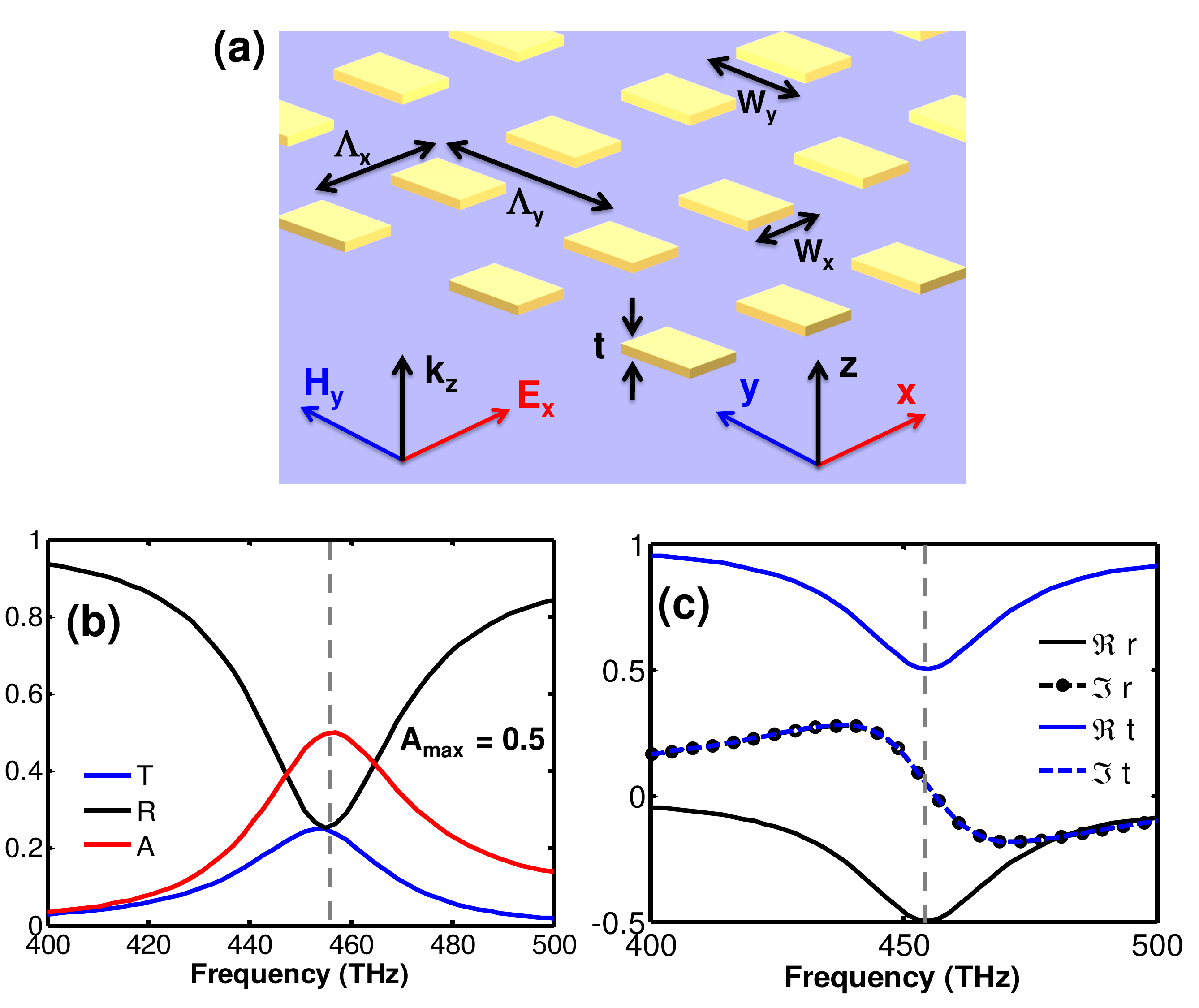}
\caption{(a) Schematic view of the rectangular nanopatches designed for maximum
absorption, i.e. $A_{\textrm{max}}=\frac{1}{2}$. (b) The reflection,
transmission and absorption spectra. (c) The reflection and transmission
coefficients. Note that at resonance, $r|_{\textrm{res}}=-\frac{1}{2}$
and $t|_{\textrm{res}}=\frac{1}{2}$. The period in x and y directions
are $\Lambda_{x}=320$~nm and $\Lambda_{y}=500$~nm.
The gold nanopatch has a thickness of $t=15$ nm and a width of $\textrm{W}_{x}=100$~nm
and $\textrm{W}_{y}=145$~nm. \label{fig:TR_SingleNanopaches_50_Absorption}}
\end{figure}

As a practical example, we investigate an array of gold nanopatches
which exhibits only an electric dipolar response. It is possible to achieve
the maximal absorption for the array by proper varying the geometrical
parameters of the nanopatches and the periodicity such that it satisfies
Eq.~\ref{eq:Alpha_ED_A_Max}. This is shown in Fig.~\ref{fig:TR_SingleNanopaches_50_Absorption}~(b).
Note that the results [Figs.~\ref{fig:TR_SingleNanopaches_50_Absorption} (b) and (c)]
are in good agreement with the analytical findings explained in the
previous chapter [Figs.~\ref{fig:-UniLimitation_EDipole} (d) and (e)].

Finally, due to the duality, the response of an array of magnetic dipoles is similar to
the electric one. Therefore, it is important to highlight that an array of magnetic dipoles maximally absorbs $50$ percent of impinging light.

We have so far presented a summary of maximum possible absorption of an electrically or a magnetically resonant array of dipole nanoantennas. However, there are several approaches to increase the absorption in such arrays which, in some cases, even allows to achieve full absorption. One method is to take the advantage of dipole nanoantennas that simultaneously support both electric and magnetic responses, as discussed in Refs.~\onlinecite{Landy:08,Alaee:15,yazdinew,Radi2015}. In the next section, we theoretically review this  problem.
\section{Optical properties of an array of electric and magnetic nanoantennas}
\begin{figure}
\begin{centering}
 \includegraphics[width=0.8\columnwidth]{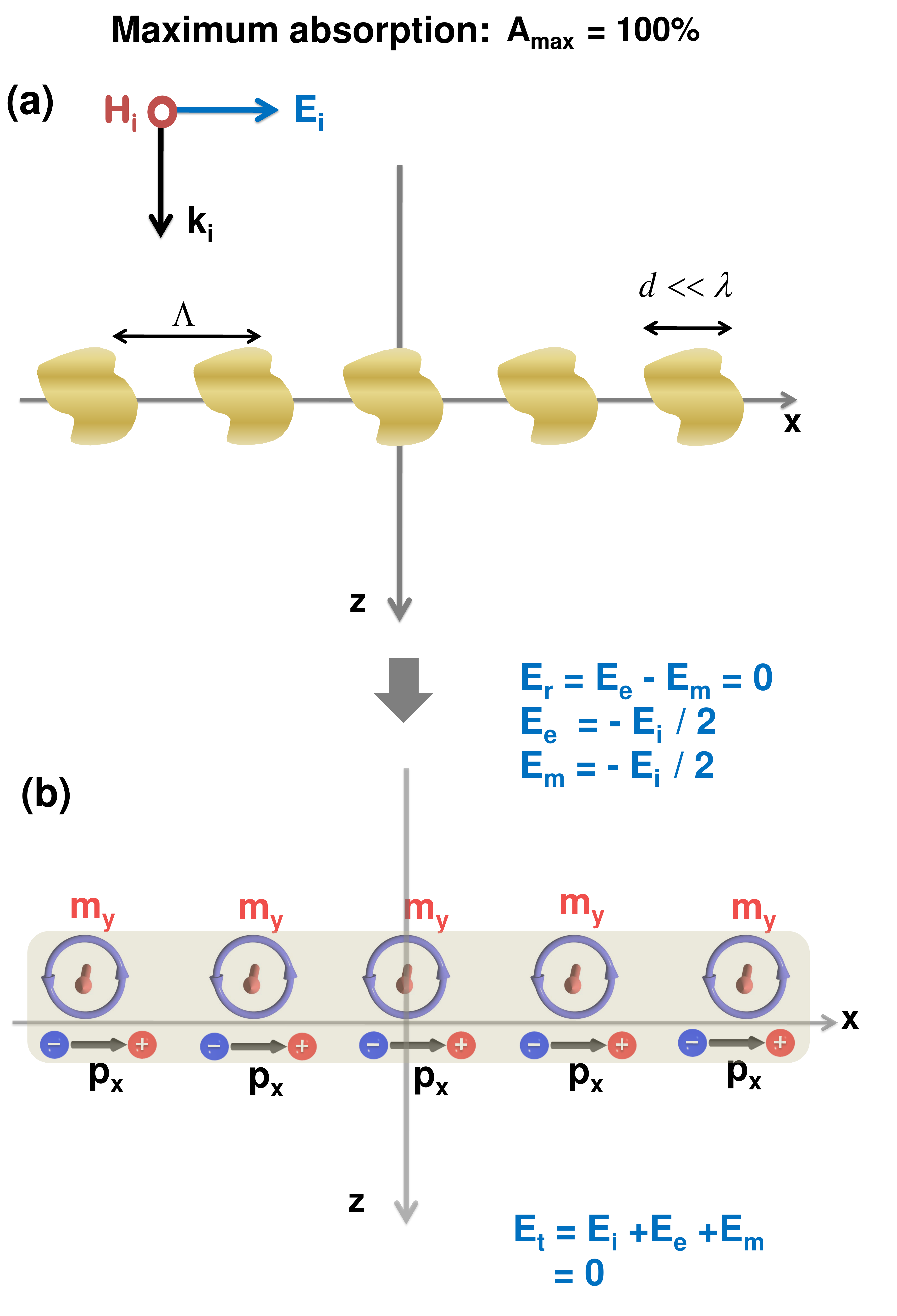}
\par\end{centering}
\caption{(a) Geometry of an infinite periodic array of nanoantennas that support
electric and magnetic dipole moments in a uniform host medium excited
by a plane wave. The two-dimensional array is periodic in x-y plane.
(b) show the radiated field by the average magnetic current density
in the case of maximum absorption.\label{fig:Array_E_M_Dipoles} }
\end{figure}

Let us assume that an array of electric and magnetic dipole particles is located in
the $\textrm{xy}$-plane and is illuminated by a plane wave with an electric
field polarized along the $x$-axis that propagates in the positive $z$-direction (Fig.~\ref{fig:Array_E_M_Dipoles}). The period of the array is $\Lambda$, which is considered to be sufficiently small compared to the wavelength $\lambda$. The constitutive relations between the induced dipole moments ($p_{x},m_{y}$) and local fields ($E_{x}^{\textrm{loc}}$,$H_{y}^{\textrm{loc}}$)
read\begin{eqnarray*}
p_{x} & = & \epsilon_{0}\alpha_{\textrm{ee}}E_{x}^{\textrm{loc}},\\
m_{y} & = & \alpha_{mm}H_{y}^{\textrm{loc}},
\end{eqnarray*}
where $E_{x}^{\textrm{loc}}=E_{x}^{\textrm{inc}}+E_{x}^{\textrm{int}}$
and $H_{y}^{\textrm{loc}}=H_{y}^{\textrm{inc}}+H_{y}^{\textrm{int}}$.
By using the generalized boundary conditions and the averaged electric
and magnetic current densities~\cite{newIdem, Holloway:05,Tretyakov:03,Albooyeh:2015,MohammadA}, we write the reflected $E_{x}^{\textrm{r}}$
and transmitted $E_{x}^{\textrm{t}}$ electric fields in terms of the
electric and magnetic dipole moments as\begin{eqnarray}
E_{x}^{\textrm{r}} & = & \frac{i\omega}{2\Lambda^{2}}\left(Z_{0}p_{x}-\mu_{0}m_{y}\right),\label{eq:Er_EM}\\
E_{x}^{\textrm{t}} & = & E_{x}^{\textrm{inc}}+\frac{i\omega}{2\Lambda^{2}}\left(Z_{0}p_{x}+\mu_{0}m_{y}\right).\label{eq:Et_EM}
\end{eqnarray}

Next, by employing the relation between the local and incident fields, we obtain the dipole moments in terms of the incident
fields, i.e., \begin{eqnarray}
p_{x} & = & \epsilon_{0}\frac{\alpha_{\textrm{ee}}}{1-\alpha_{\textrm{ee}}\beta_{\textrm{ee}}}E_{x}^{\textrm{inc}}=\epsilon_{0}\alpha_{\textrm{ee}}^{\textrm{eff}}E_{x}^{\textrm{inc}},\label{eq:px_EM}\\
m_{y} & = & \frac{\alpha_{\textrm{mm}}}{1-\alpha_{\textrm{mm}}\beta_{\textrm{mm}}}H_{y}^{\textrm{inc}}=\alpha_{\textrm{mm}}^{\textrm{eff}}H_{y}^{\textrm{inc}},\label{eq:my_EM}
\end{eqnarray}
where $\alpha_{\textrm{ee}}^{\textrm{eff}}$ and $\alpha_{\textrm{mm}}^{\textrm{eff}}$
are referred as the effective electric and magnetic polarizabilities of the array. By substituting Eqs.~\ref{eq:px_EM}
and \ref{eq:my_EM} into Eq.~\ref{eq:Er_EM} and Eq.~\ref{eq:Et_EM}, the reflection and transmission coefficients of the array read
\begin{eqnarray}
r & = & \frac{E_{x}^{\textrm{r}}}{E_{x}^{\textrm{i}}}=\frac{ik}{2\Lambda^{2}\Delta_{\textrm{ee}}\Delta_{\textrm{mm}}}\left(\alpha_{\textrm{ee}}-\alpha_{\textrm{mm}}\right),\label{eq:R_EM}\\
\nonumber t & = & \frac{E_{x}^{\textrm{t}}}{E_{x}^{\textrm{i}}}=1+\frac{ik}{2\Lambda^{2}\Delta_{\textrm{ee}}\Delta_{\textrm{mm}}}\left[\alpha_{\textrm{ee}}+\alpha_{\textrm{mm}}\right.\\
 &  & \left.-\alpha_{\textrm{ee}}\alpha_{\textrm{mm}}\left(\beta_{\textrm{ee}}+\beta_{\textrm{mm}}\right)\right],
\label{eq:T_EM}
\end{eqnarray}
where $\Delta_{\textrm{ee}}=1-\alpha_{\textrm{ee}}\beta_{\textrm{ee}}$
and $\Delta_{\textrm{mm}}=1-\alpha_{\textrm{mm}}\beta_{\textrm{mm}}.$
Note that $\beta_{\textrm{ee}}=\beta_{\textrm{mm}}$. Finally, the
absorption is found as~\cite{Tretyakov:03,Albooyeh:12,Alaee:15}\begin{eqnarray}
A & = & 1-\left|r\right|^{2}-\left|t\right|^{2}\nonumber \\
 & = & 1-\left|r_{\textrm{ee}}-r_{\textrm{mm}}\right|^{2}-\left|1+r_{\textrm{ee}}+r_{\textrm{mm}}\right|^{2},
\end{eqnarray}
where $r_{\textrm{ee}}=\frac{ik}{2\Lambda^{2}}\alpha_{\textrm{ee}}^{\textrm{eff}}$
and $r_{\textrm{mm}}=\frac{ik}{2\Lambda^{2}}\alpha_{\textrm{mm}}^{\textrm{eff}}$.
It can be shown that the maximum absorption occurs when \begin{equation}
r_{\textrm{ee}}=r_{\textrm{mm}}=-\frac{1}{2}\,\,\,\Rightarrow\,\,\,\alpha_{\textrm{ee}}^{\textrm{eff}}=\alpha_{\textrm{mm}}^{\textrm{eff}}.
\end{equation}

The above condition, i.e. $\alpha_{\textrm{ee}}^{\textrm{eff}}=\alpha_{\textrm{mm}}^{\textrm{eff}}$, is called balanced (Kerker) condition for the effective polarizabilities. Hence, the balanced condition for the individual polarizabilities
is given by\e
{\alpha}_{\textrm{ee}}  =  {\alpha}_{\textrm{mm}}=\frac{1}{\beta_{\mathrm{ee}}-\frac{ik}{\Lambda^{2}}}.
\f
\begin{figure}
\begin{centering}
\includegraphics[width=0.98\columnwidth]{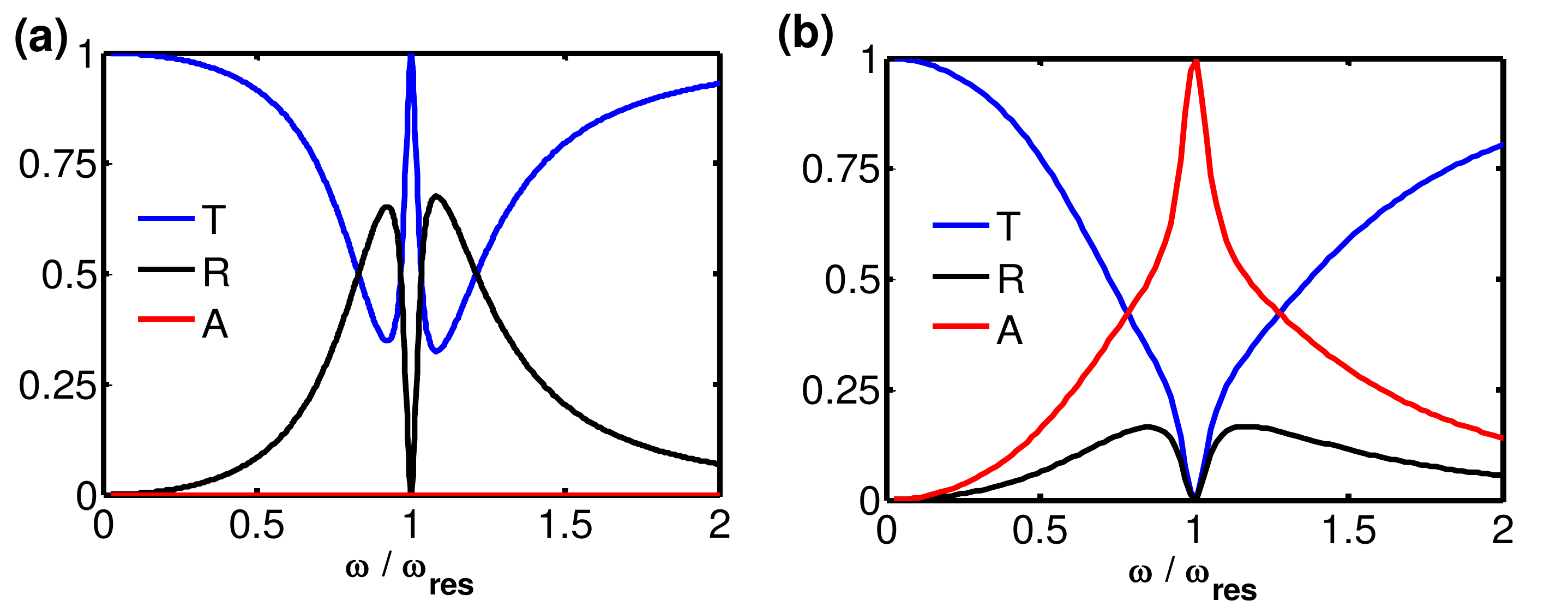}
\par\end{centering}

\caption{(a) The transmission, reflection, and absorption as a function of
normalized frequency for the lossless nanoantennas, i.e. $\gamma_{\textrm{ee}}=0$
and $\gamma_{\textrm{mm}}=0$. (b) The same plot as (a) for the complete
absorption case that requires, i.e. $\gamma_{\textrm{ee}}=\frac{\alpha_{\textrm{0ee}}}{2\Lambda^{2}c_{0}}$
and $\gamma_{\textrm{mm}}=\frac{\alpha_{\textrm{0mm}}}{2\Lambda^{2}c_{0}}$.
We assumed that the $\Lambda=\lambda_{\textrm{res}}/5$ and $\alpha_{\textrm{0ee}}=8\alpha_{\textrm{0mm}}=\Lambda\times c_{0}^{2}$.
\label{fig:Max_ED_MD}}
\end{figure}

Note that the maximum absorption for an array of electric and magnetic
dipoles (i.e.~$A_{\textrm{max}}=1$) is twice the absorption that
can be obtained from an array of only electric or magnetic dipoles. At the maximum
absorption, the array will completely cancel the incident field in the forward direction without generating any backward wave, i.e. reflection {[}Fig.~\ref{fig:Array_E_M_Dipoles} (b){]}. Moreover, the relation between the induced averaged electric and magnetic surface current
densities in this case reads
\begin{eqnarray}
\frac{J_{y}^{\textrm{m}}}{J_{x}^{\textrm{e}}} & = & \frac{E_{x}^{\textrm{inc}}}{E_{x}^{\textrm{inc}}/Z_{0}}=Z_{0},
\end{eqnarray}
which simply means that the array is impedance matched with the embedding
medium. The relation between electric and magnetic dipole moments of each nanoantenna
in the total absorption regime, i.e. the so-called Kerker condition, is $p_{\textrm{x}}=m_{\textrm{y}}/c$.

In to order to understand the mechanism of total absorption for an array with both electric and magnetic dipole moments, let us assume
that the dispersive characteristics of the magnetic polarizability of the nanoantenna can be described by a Lorentzian line-shape (see Eq.~\ref{eq:Alpha_ED} for the electric polarizability) that read as~\cite{Sipe:74,Albooyeh:12}
\begin{eqnarray}
\frac{1}{\alpha_{\textrm{mm}}} & = & \frac{\omega_{\textrm{0mm}}^{2}-\omega^{2}-i\omega\gamma_{\textrm{mm}}}{\alpha_{\textrm{0mm}}}-i\frac{k^{3}}{6\pi},
\end{eqnarray}
where $\omega_{\textrm{0mm}}$ is the magnetic resonance angular frequency, $\gamma_{\textrm{0mm}}$ is the Ohmic losses and $\alpha_{\textrm{0mm}}$ is the corresponding oscillator amplitude.
By using Eq.~\ref{eq:R_ED} and the inter-particle interaction constant,
i.e. $\beta_{\textrm{ee}}=\beta_{\textrm{mm}}=\Re\left(\beta_{\textrm{ee}}\right)-i\frac{k^{3}}{6\pi}+i\frac{k}{2\Lambda^{2}}$,
the reflection coefficient is given by\begin{eqnarray}
\nonumber r & = & \frac{ik}{2\Lambda^{2}}\left(\frac{1}{\frac{1}{\alpha_{\textrm{ee}}}-\beta_{\textrm{ee}}}-\frac{1}{\frac{1}{\alpha_{\textrm{mm}}}-\beta_{\textrm{mm}}}\right)\\
  & = & \frac{ik}{2\Lambda^{2}}\left[\frac{\alpha_{\textrm{0ee}}}{\omega_{\textrm{0ee}}^{'2}-\omega^{2}-i\omega\left(\gamma_{\textrm{ee}}+\frac{\alpha_{\textrm{0ee}}}{2\Lambda^{2} c_{0}}\right)}\right.\nonumber \\
 &  & \left.-\frac{\alpha_{\textrm{0mm}}}{\omega_{\textrm{0mm}}^{'2}-\omega^{2}-i\omega\left(\gamma_{\textrm{mm}}+\frac{\alpha_{\textrm{0mm}}}{2\Lambda^{2}c_{0}}\right)}\right]\nonumber \\
 & = & \frac{ik}{2\Lambda^{2}}\left(\alpha_{\textrm{ee}}^{\textrm{eff}}-\alpha_{\textrm{mm}}^{\textrm{eff}}\right),
\end{eqnarray}
where $\omega_{\textrm{0mm}}^{'}=\omega_{\textrm{0mm}}-\Re\left(\beta_{\textrm{mm}}\right)\alpha_{\textrm{0mm}}$
and $\omega_{\textrm{0ee}}^{'}=\omega_{\textrm{0ee}}-\Re\left(\beta_{\textrm{ee}}\right)\alpha_{\textrm{0ee}}$
are the resonance frequencies of the electric and magnetic dipoles in the array, respectively. The transmission coefficient correspondingly reads as \begin{eqnarray*}
t & = & 1+\frac{ik}{2\Lambda^{2}}\left(\alpha_{\textrm{ee}}^{\textrm{eff}}+\alpha_{\textrm{mm}}^{\textrm{eff}}\right).
\end{eqnarray*}

Let us first consider a lossless metasurface. We assume that
$\omega_{\textrm{0ee}}^{'}=\omega_{\textrm{0mm}}^{'}=\omega_{\textrm{res}}$. When the dipole nanoantennas in the array are lossless, i.e. $\gamma_{\textrm{ee}}=\gamma_{\textrm{mm}}=0$, the transmission, reflection, and absorption as a function of normalized frequency are shown in Fig.~\ref{fig:Max_ED_MD}~(a).
In this case, the transmission possesses similar features to structures known
in the context of electromagnetically induced transparency (EIT)~\cite{Yang:14}.
The array shows zero absorption and the reflection vanishes at the resonance frequency.

Next, we consider the case of maximum absorption (known as critical coupling). In this case, the Ohmic losses for the electric and magnetic dipoles are $\gamma_{\textrm{ee}}=\frac{\alpha_{\textrm{0ee}}}{2\Lambda^{2}c_{0}}$,
$\gamma_{\textrm{mm}}=\frac{\alpha_{\textrm{0mm}}}{2\Lambda^{2}c_{0}}$,
respectively. Figure~\ref{fig:Max_ED_MD}~(b) depicts the spectral
dependency of the optical coefficients in such situation. At the resonance
frequency, we observe total absorption. Indeed, one achieves the electromagnetically induced absorption (EIA) by proper tuning of the electric and magnetic dipole moments~\cite{Taubert:12}. Several geometries have been proposed to achieve total absorption based on balanced electric and magnetic moments~\cite{Landy:08,Watts:12,Alaee:13,Alaee:15,yazdinew,Radi2015,AlbooyehAbsorber}.

Besides the discussed scenario to obtain the total absorption in an array of dipole particles, there are other alternatives which exploit metasurfaces with either electric or magnetic responses while taking the advantage of a back reflector in their design which is the topic of the rest of this review study.

Next, we focus on this alternative approach to enhance the absorption of an array of electric dipole moment on top of a reflecting ground plane. As we will see, this alternative also allows to fully suppress the transmission while canceling out the reflection at the same time.

\section{Asymmetric Fabry-Perot cavity and perfect absorbers}

\begin{figure}
 \centering
  \includegraphics[width=0.98\columnwidth]{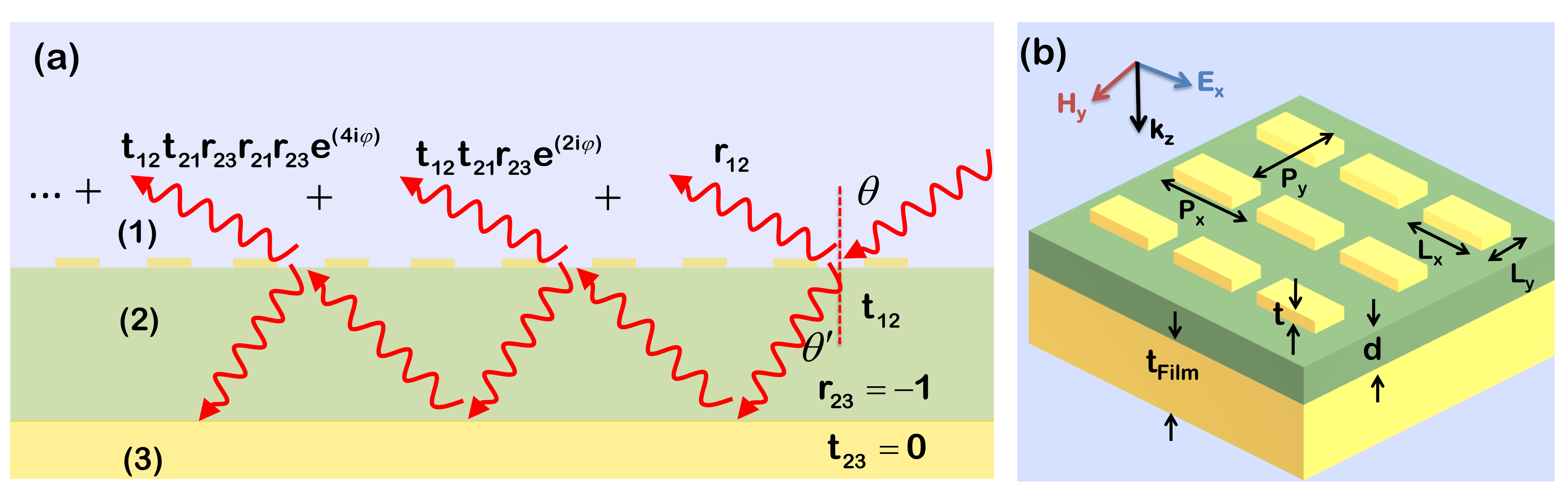}
\caption{(a) Asymmetric Fabry-Perot cavity model for a perfect absorber in
the far-field interference scheme. (b) Schematic of the investigated perfect
absorber. The geometrical parameters of the structure are: $t_{\mathrm{Film}}=300\,$
nm, $d=20-1500\,$nm, $L_{x}=400\,$nm, $L_{y}=100\,$nm, $P_{x}=P_{y}=500$~nm
and $t=20$~nm. \label{fig:FFS_Geo} }
\end{figure}

In the previous section, we have shown that the absorption of an array
of rectangular nanopatches is limited to $50$ percent. Note that this
is a universal limitation as long as the nanopatches can be described
by an electric dipole moment only, i.e. all higher order multipoles are negligible.
One of the simplest and practical approaches to enhance the absorption
is to use a metallic ground plate in order to suppress the transmission
{[}Fig.~\ref{fig:FFS_Geo}~(a){]}. In fact, complete light absorption
occurs when the reflection goes to zero ($R=0$)  since the transmission is zero ($T=0$) (due to the thick metallic ground plate).
In order to achieve zero reflection, the dielectric spacer that separates the nanopatches from the ground plate should be
properly tuned such that the directly reflected light at the nanopatch array interferes destructively
with the light that experiences multiple reflections in the layered system~{[}Fig.~\ref{fig:FFS_Geo}~(a){]}.
In the following, we will show that the optical response of the structure
can be fully explained by a simple Fabry-Perot model. This model is
only valid for considerably thick dielectric spacers when the effects of near-field coupling between ground plate and nanopatch array is negligible.

\subsection{Theory of ground-back perfect absorption of electrically resonant sheets}
The investigated perfect absorber consists of a metasurface
on top of a metallic ground plate separated by a thick dielectric
spacer~{[}Fig.~\ref{fig:FFS_Geo} (a){]}. To simplify the structure
and also to provide a physical explanation, we consider the structure
as an asymmetric Fabry-Perot cavity with two mirrors, i.e., the metallic
ground plate as the bottom mirror and the array of nanoantennas as
the top mirror. By using Airy's formula, the total reflection and transmission coefficients read~\cite{Saleh:07,Alaee:12,Chen:12}:
\begin{equation}
r=r_{\mathrm{12}}+r_{\mathrm{m}}=r_{\mathrm{12}}+\frac{t_{\mathrm{12}}t_{\mathrm{21}}r_{\mathrm{23}}e^{(2i\varphi)}}{1-r_{\mathrm{21}}r_{\mathrm{23}}e^{(2i\varphi)}},\label{EQ_Airy}
\end{equation}

\begin{equation}
t=\frac{t_{\mathrm{12}}t_{\mathrm{23}}e^{(2i\varphi)}}{1-r_{\mathrm{21}}r_{\mathrm{23}}e^{(2i\varphi)}},\label{EQ_Airy-1}
\end{equation}
where $t_{12}$, $t_{21}$, $r_{12}$, $r_{21}$, $r_{23}$ are complex-valued
reflection and transmission coefficients at both interfaces, $\varphi=k_{0}nd\cos\theta^{\prime}$
is the phase accumulated upon a single cavity transfer, $k_{0}$ is the free
space wavenumber, $n$ is the refractive index of the dielectric spacer,
and $d$ is its thickness. A full wave simulation is used to calculate the reflection and
transmission coefficients ($r_{12}$, $t_{12}$, $r_{21}$, and $t_{21}$) where the array of nanoantennas is assumed to be sandwiched between two semi-infinite half-spaces, i.e. air and dielectric and is illuminated by a linearly polarized incident plane wave either from the top air or from the bottom dielectric~\cite{Alaee:12}.

As already mentioned, a perfect absorption is achievable if the
reflection as well as transmission of the structure vanishes. The
transmission of the structure is totally suppressed in any case, i.e. $T=|t|=0$ due to sufficiently thick metallic ground plate (i.e. $t_{23}=0$).
To achieve total absorption ($A=1-T-R\simeq1$) the reflection should
be simultaneously zero ($R=|r|^{2}=|r_{12}+r_{m}|^{2}\simeq0$). The reflection
is the sum of the direct reflection coefficient $r_{12}$ and the
multiple reflection coefficient $r_{m}$ {[}as sketched in Fig. \ref{fig:FFS_Geo}~(a){]}.
Therefore, in the far-field scheme, the condition of the perfect absorption
reads

\begin{equation}
\mbox{\ensuremath{r_{12}=-r_{m}.}}\label{eq:PA_FarField}
\end{equation}

In the next subsection, we will show that the total absorption occurs
whenever Eq.~\ref{eq:PA_FarField} is fulfilled.

\begin{figure}
 \centering
  \includegraphics[width=0.98\columnwidth]{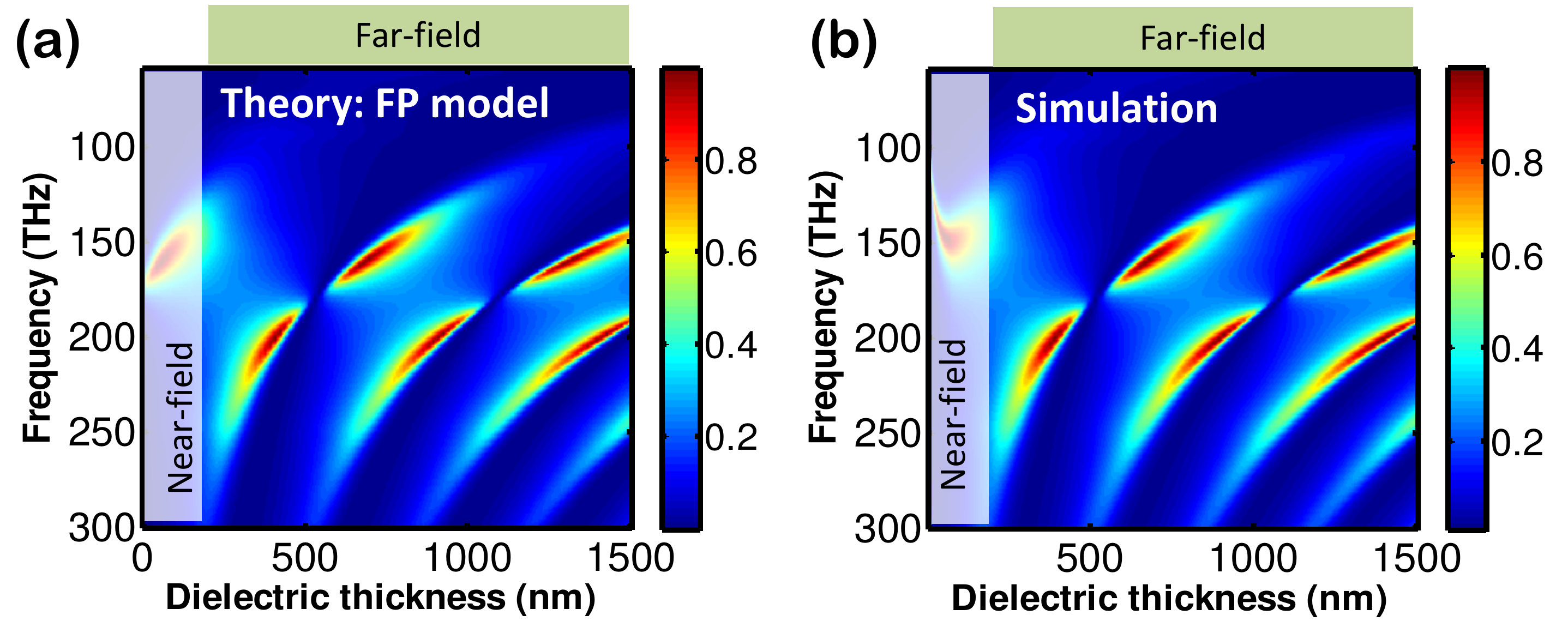}
\caption{(a) and (b) Theoretical and simulated absorption spectra as a function
of the frequency and the dielectric thickness ($d$) of the spacer for a gold nanopatch based
perfect absorber at normal incidence. \label{fig:FFS_Absorption}}
\label{FIG_GPA-1}
\end{figure}

\subsection{Array of nanopatches on top of a metallic ground plate }

The schematic view of the investigated perfect absorber is depicted in
Fig.~\ref{fig:FFS_Geo}~(b). It consists of an array of rectangular
gold nanopatches on top of a metallic ground plate separated by a
thick dielectric spacer. The structure is periodic in $x$ and $y$
directions and the periods are $P_{x}=P_{y}=500$~nm. The refractive
index of the dielectric spacer is $n=\sqrt{\varepsilon_{\mathrm{d}}}=1.5$. To explore the underlying physics of the investigated perfect absorber,
we calculated the absorption \emph{A} as a function of the thickness
of the dielectric spacer $d$ and the frequency {[}Fig.~\ref{fig:FFS_Absorption}~(b){]}.
The dielectric spacer is varied from $20-1500$~nm. To be able to
compare it with the Fabry-Perot model, the simulated result is divided
into two distinct parts, namely, far-field and near-field~{[}Fig.~\ref{fig:FFS_Absorption}~(a)
and (b){]}. The numerical result shows that the perfect absorption
($A=1$) can be achieved for various dielectric spacer thicknesses $d$, periodically
spaced around a frequency of operation. Moreover, the resonance frequency
of maximum absorption is slightly changing for different dielectric
spacer. Note that the hybridization of the localized eigenmode of
the gold nanopatches with the Fabry-Perot resonance of the cavity
can be clearly seen in Fig.~\ref{fig:FFS_Absorption}~(a) and (b).
This leads to a strong Rabi splitting and avoid crossing {[}Fig.~\ref{fig:FFS_Absorption}~(a)
and (b){]}.

\begin{figure}
 \centering
  \includegraphics[width=0.98\columnwidth]{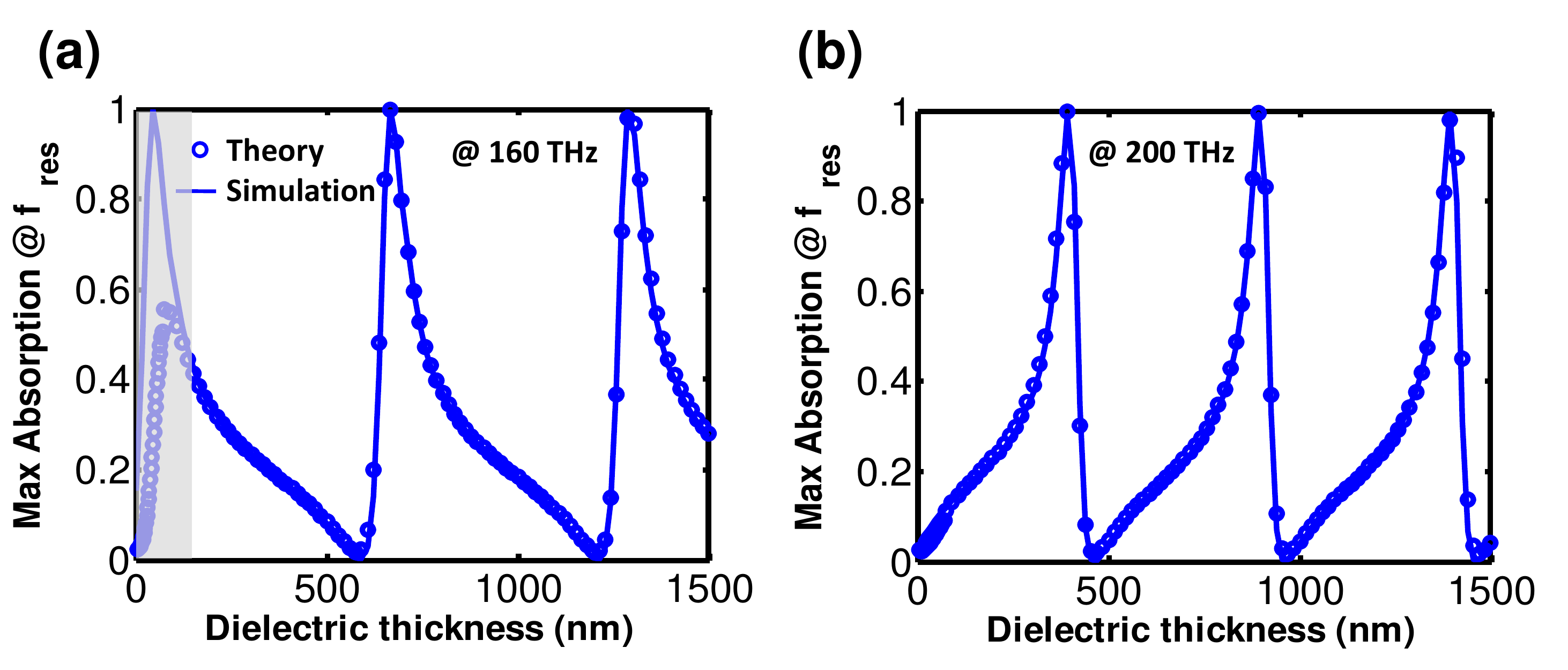}
\caption{Maximum absorption as a function of the thickness of the dielectric spacer
calculated from the semi-analytical approach and from full wave simulations
for two different frequencies at $160$~THz (a) and at $200$~THz (b), respectively.
\label{fig:FFS_SemiAn_Numerical} }
\end{figure}
The absorption spectra as a function of the dielectric spacer thickness
$d$ and resonance frequency is calculated by using the Airy's formula which shows similar behavior compared to the simulated
results~{[}Fig.~\ref{fig:FFS_Absorption}~(a) and (b){]}. As already
highlighted, the analytical result does not predict the simulated result
for thin dielectric spacers due to the strong coupling between nanopatches
and the metallic ground plate. This can be better seen in Fig.~\ref{fig:FFS_SemiAn_Numerical}
for two different resonance frequencies $160$~THz (a) and $200$~THz
(b). The absorption is shown at those discrete
frequencies as a function of the thickness of the dielectric spacer.
\begin{figure}
 \centering
  \includegraphics[width=0.98\columnwidth]{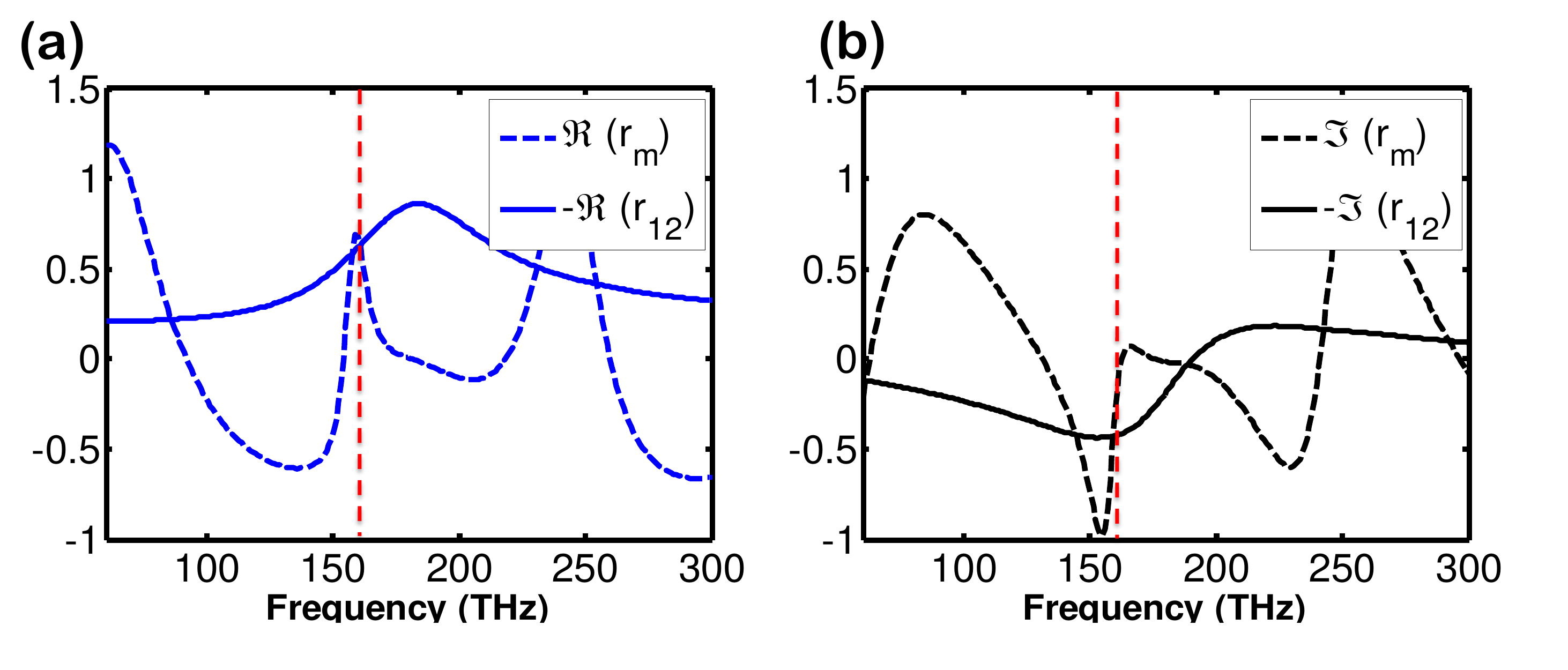}
\caption{(a) and (b) Real and imaginary parts of direct reflection coefficient
($r_{12}$) as well as multiple reflection coefficient ($r_{m}$)
with dielectric thickness $d=370\,$nm. The red dashed lines represent
the crossing point when real and imaginary parts of reflection coefficients
have the same magnitude but opposite sign, i.e. $r_{12}=-r_{m}$.\label{fig:FFS_R_cofficients} }
\end{figure}

The analytical and numerical results are displayed in the same figure {[}Fig.\,\ref{fig:FFS_SemiAn_Numerical}~(a)
and (b){]} for comparison. For thick dielectric spacers, i.e. $d>250$~nm, there
is a perfect agreement between the numerical and analytical
results. However, for thin dielectric spacers, there is a huge deviation due to strong
coupling between the metallic ground plate and the nanopatches. This has been investigated in detail in Refs.~\onlinecite{Ciraci:12,Alaee:13}.
In order to fully understand the absorption mechanism in the investigated
structure, the real and imaginary parts of the direct reflection coefficient
$r_{12}$ and the multiple reflection coefficient $r_{m}$ are calculated
at the resonance frequency $f=160$~THz for a spacer thickness $d=370$~nm
{[}Fig.~\ref{fig:FFS_R_cofficients}~(a) and \ref{fig:FFS_R_cofficients}~(b){]}.
The total reflection coefficient is completely suppressed, i.e. $r=r_{12}+r_{m}=0.$, due to the destructive interference between both the reflection coefficients $r_{m}$ and $r_{12}$, i.e. $r_{12}=-r_{m}$.  Therefore, the absorption reaches nearly $100$ percent for the investigated frequency.

\section{Conclusions}

We have presented a review where by analytical means the universal limitation on the
scattering/absorption cross section of a dipolar nanoantenna that supports an electric and/or a magnetic response was studied. We have provided a concise theoretical review that shows that a metasurface made of nanoantennas that sustain only either electric or magnetic dipole moments maximally absorbs $50$ percent of the impinging light. To overcome this
limitation, we have theoretically revisited two known alternatives to obtain a perfect absorber. One of them was established on the use of dipolar nanoantennas that simultaneously support both electric and magnetic responses whereas the other was based on a periodic array of electrically/magnetically resonant nanoantennas on top of a reflector. Moreover, for the second alternative, we have applied a simple semi-analytical approach based on an asymmetric Fabry-Perot model in order to investigate the optical responses of such a perfect absorber. The underlying mechanism of the total absorption is fully explained by destructive interference. We have provided a laconic explanation of underlying mechanism of the maximum absorption in each study case.

By providing a useful tutorial, the goal of the present work has been to bring about a correct physical insight in understanding the scattering/absorption phenomena both in the level of an individual meta-atom and also in the level of many inclusions of meta-atoms composing a metasurface. We believe that the current review would be helpful to a wide range of audience both in the junior and senior levels due to the concise presentation manner.\\

\section{Acknowledgments}
R.A. would like to acknowledge financial support from the Max Planck Society.

\appendix
\section{Derivation of scattered and extinct power for an electric dipole}

Here we present a derivation for the scattered and extinct power shown in Eqs. (\ref{eq:P_scat}) and (\ref{eq:Cext_ED})  for an electrically dipolar polarizable particle when excited by an electromagnetic field. We assume that the particle contains only electric dipole moment, while similar expression may be derived for a magnetic dipole moment.

From the Poynting theorem~\cite{Jackson:98}, we have
\begin{equation}
\begin{array}{l}
\displaystyle
\int_{V}\nabla \cdot \left( \mathcal{E}_{\rm tot}\times\mathcal{H}_{\rm tot} \right) dv =-\int_{V}\mathcal{J}\cdot \mathcal{E}_{\rm tot} dv
\vspace*{.2cm}\\\displaystyle
\hspace*{3.8cm}
-\int_{V}\mathcal{E}_{\rm tot}\cdot {\partial\mathcal{D}\over\partial t} dv
\vspace*{.2cm}\\\displaystyle
\hspace*{3.8cm}
-\int_{V}\mathcal{H}_{\rm tot}\cdot {\partial\mathcal{B}\over\partial t} dv,
\end{array}\l{Poynting1}
\end{equation}
where $\mathcal{E}_{\rm tot}$ and $\mathcal{H}_{\rm tot}$ are the time varying {\it total} electric and magneic fields while $\mathcal{D}$ and $\mathcal{B}$ are the electric displacement vector and magnetic induction, respectively, and $\mathcal{J}$ is a current distribution in volume $V$. Considering the time harmonic fields with the time dependence ${\rm e}^{-i\omega t}$ (i.e., the relation between the time harmonic fields is considered to be $\mathcal{A}=\Re\{\-A {\rm e}^{-i\omega t}\}$), the Poynting theorem reads (after averaging over a time cycle)
\begin{equation}
\begin{array}{l}
\displaystyle
\int_{V}\nabla \cdot {1\over 2}\Re\{ \-E_{\rm tot}\times\-H_{\rm tot}^\ast \} dv =-\int_{V} {1\over 2}{\rm Re}\{\-J^\ast\cdot \-E_{\rm tot}\} dv
\vspace*{.2cm}\\\displaystyle
\hspace*{4.2cm}
-{\omega\over 2}\int_{V}{\rm Im} \{\-E_{\rm tot}^\ast\cdot \-D\} dv
\vspace*{.2cm}\\\displaystyle
\hspace*{4.2cm}
-{\omega\over 2}\int_{V}{\rm Im} \{\-H_{\rm tot}^\ast\cdot \-B\} dv.
\end{array}\l{PoynTime}
\end{equation}
Considering a non-conductive, non-magnetic particle with only electrical properties; we have $\-J^\ast \cdot \-E_{\rm tot}=0$, $\-D=\epsilon_0\-E_{\rm tot}+\-P$, and $\-B=\mu_0\-H_{\rm tot}$. Substituting these quantities into~\r{PoynTime}, we reach at
\e
{1\over 2}\int_{V}\nabla \cdot{\rm Re}\{ \-E_{\rm tot}\times\-H_{\rm tot}^\ast \} dv ={\omega\over 2}\int_{V}{\rm Im} \{\-P^\ast\cdot \-E_{\rm tot}\} dv. \l{PoynTime1}
\f
The left hand side of Eq.~\r{PoynTime1} is the total power going out of a surface embedding volume $V$ which contains the particle, while the right hand side is the negative of the absorbed power since the absorbed power is going into the surface $S$ of volume $V$ (the work done by the field on the dipolar particle). Therefore, the absorbed power by one dipolar particle reads
\e
P_{\rm abs} =-{\omega\over 2}{\rm Im} \{\-p^\ast\cdot \-E_{\rm tot} \}, \l{ABS}
\f
since $\-P=\int\-pdv$. The electric field in~\r{ABS} is the total field at the position of the particle, which is the contribution of the local field and the scattered field by the dipole; i.e,
\e
\-E_{\rm tot}=\-E_{\rm loc}+\-E_{\rm sca}.\l{tot}
\f
Since we have one particle, then the local field is also the excitation field; i.e., $\-E_{\rm loc}=\-E_{\rm inc}$, and plugging~\r{tot} into~\r{ABS} the absorbed power reads
\begin{equation}
\begin{array}{l}
\displaystyle
P_{\rm abs} =-{\omega\over 2}{\rm Im} \{\-p^\ast\cdot \-E_{\rm tot} \}
\vspace*{.2cm}\\\displaystyle
\hspace*{.75cm}
=-{\omega\over 2}{\rm Im} \{\-p^\ast\cdot \-E_{\rm inc} \}-{\omega\over 2}{\rm Im} \{\-p^\ast\cdot \-E_{\rm sca} \}.
\end{array}\l{ABSEXT}
\end{equation}
Let us now calculate the second part of the right hand side of Eq.~\r{ABSEXT}, i.e., ${\omega\over 2}{\rm Im} \{\-p^\ast\cdot \-E_{\rm sca}\}$. The scattered electric field by a dipole reads~\cite{Jackson:98}
\begin{equation}
\begin{array}{l}
\displaystyle
\-E_{\rm sca}={{\rm e}^{ikr}\over 4\pi\epsilon_0} \left[ \left(\-n\times\-p\right)\times \-n {k^2\over r}
\right.\vspace*{.2cm}\\\displaystyle
\hspace*{.8cm}\left.
+\left[3\-n\left(\-n\cdot\-p\right)-\-p\right]\left({1\over r^3}-{ik\over r^2}\right) \right],
\end{array}\l{dipoleScat}
\end{equation}
then, we inner product both sides by $\-p^\ast$ to obtain
\begin{equation}
\begin{array}{l}
\displaystyle
\-p^\ast\cdot \-E_{\rm sca}={{\rm e}^{ikr}\over 4\pi\epsilon_0} \left[ \-p^\ast\cdot\left[\left(\-n\times\-p\right)\times \-n\right] {k^2\over r}
\right.\vspace*{.2cm}\\\displaystyle
\hspace*{.8cm}\left.
+\left[3\left(\-n\cdot\-p^\ast\right)\left(\-n\cdot\-p\right)-|\-p|^2\right]\left({1\over r^3}-{ik\over r^2}\right) \right].
\end{array}\l{eqT}
\end{equation}
Next, if we use the identity $
\-p^\ast\cdot\left[\left(\-n\times\-p\right)\times \-n\right]=|\-p|^2-\left(\-n\cdot\-p^\ast\right)\left(\-n\cdot\-p\right)
$ in Eq.~\r{eqT}, the result reads
\begin{equation}
\begin{array}{l}
\displaystyle
\-p^\ast\cdot \-E_{\rm sca}={{\rm e}^{ikr}\over 4\pi\epsilon_0} \left( \left[|\-p|^2-\left(\-n\cdot\-p^\ast\right)\left(\-n\cdot\-p\right)\right] {k^2\over r}
\right.\vspace*{.2cm}\\\displaystyle
\hspace*{.8cm}\left.
+\left[3\left(\-n\cdot\-p^\ast\right)\left(\-n\cdot\-p\right)-|\-p|^2\right]\left({1\over r^3}-{ik\over r^2}\right) \right).
\end{array}\l{eqT1}
\end{equation}
Now, we expand the exponential term ${\rm e}^{ikr}=1+ikr-k^2r^/2-ik^3r^3/6+\cdots$ in Eq.~\r{eqT1}, and take the imaginary parts of both sides and then take their limits at $r \rightarrow 0$, (i.e., at the position of the particle). The result reads
\begin{equation}
\begin{array}{l}
\displaystyle
{\rm Im} \{\-p^\ast\cdot \-E_{\rm sca} \}\rfloor_{r\rightarrow 0}={k^3\over 4\pi\epsilon_0} \left( \left[|\-p|^2-\left( \-n\cdot\-p^\ast\right)\left(\-n\cdot\-p\right)\right]
\right.\vspace*{.2cm}\\\displaystyle
\hspace*{.8cm}\left.
+{1\over 3}\left[3\left(\-n\cdot\-p^\ast\right)\left(\-n\cdot\-p\right)-|\-p|^2\right] \right).
\end{array}\l{eqT111}
\end{equation}
and when simplified to
\e
{\rm Im} \{\-p^\ast\cdot \-E_{\rm sca} \}\rfloor_{r\rightarrow 0}={k^3\over 6\pi\epsilon_0}|\-p|^2.\l{Scat1}
\f
If we simply substitute~\r{Scat1} into~\r{ABSEXT}, we reach at
\e
P_{\rm abs} =-{\omega\over 2}{\rm Im} \{\-p^\ast\cdot \-E_{\rm inc} \}-{\omega k^3\over 12\pi\epsilon_0}|\-p|^2. \l{ABSSCA}
\f
The second term of the right hand side in~\r{ABSSCA}, equals the negative of the scattered power by an electric dipole, [see Eq. (\ref{eq:P_scat})]. Therefore, one may simply rewrite Eq.~\r{ABSEXT} as
\e
P_{\rm ext}=-{\omega\over 2}{\rm Im} \{\-p^\ast\cdot \-E_{\rm inc} \}=P_{\rm abs}+P_{\rm scat}. \l{EXTABSSCA}
\f
$P_{\rm ext}$ as discussed in Eq.~(\ref{eq:Cext_ED}) is called the extinct power and is the sum of the absorbed and the scattered powers by the dipole. Notice, due to the duality of the electric and magnetic fields/currents in the Maxwell's equations, similar formulations can be provided for a magnetic dipole with the same procedure.


\end{document}